\documentclass[letterpaper,twocolumn,10pt]{article}

\usepackage{amsmath}

\usepackage{filecontents}

\usepackage[hyphens]{url}
\usepackage{makecell}
\usepackage{listings}
\usepackage{pbox}
\usepackage{graphicx}
\usepackage{graphics}
\usepackage{xspace}
\usepackage{multicol}
\usepackage{booktabs}
\usepackage{algorithm}
\usepackage{algorithmicx}
\usepackage{algpseudocode}
\usepackage{subcaption}
\usepackage{placeins}
\usepackage{amsmath,tabstackengine}
\usepackage{csquotes}
\usepackage{hhline}
\usepackage{tikz}
\usetikzlibrary{arrows.meta}
\usepackage{booktabs}
\usepackage{pifont}
\usepackage{amssymb}
\usepackage{siunitx}
\usepackage[english]{babel}
\usepackage[utf8]{inputenc}
\usepackage[T1]{fontenc}
\usepackage{amsmath}
\usepackage{graphicx}
\usepackage{pgfplots}
\usepackage{caption}
\usepackage{paper}
\usepackage[singlelinecheck=on,font=footnotesize,labelfont=bf]{caption}
\usepackage[titletoc,title]{appendix}
\usepackage[normalem]{ulem}

\graphicspath{{figures/}}

\definecolor{dkgreen}{rgb}{0,0.6,0}
\definecolor{gray}{rgb}{0.5,0.5,0.5}
\definecolor{mauve}{rgb}{0.58,0,0.82}

\newcommand{\eg}{\textit{e.g., }}
\newcommand{\ie}{\textit{i.e., }}

\newcommand{\point}[1]{\vspace{0.8ex}\par\noindent{\bf #1:}}
\newcommand*{\rom}[1]{\textit{\expandafter\romannumeral #1}}

\newcommand{\ftodo}[1]{}

\newcommand{\pacga}{{\tt PACGA}\xspace}
\newcommand{\na}{{PTAuth}\xspace}
\newcommand{\qa}{\textit{QARMA}\xspace}
\newcommand{\ac}{{\tt AC}\xspace}

\newcommand{\pie}[1]{%
\begin{tikzpicture}
 \draw (0,0) circle (1ex);\fill (1ex,0) arc (0:#1:1ex) -- (0,0) -- cycle;
\end{tikzpicture}%
}

\stackMath

\hypersetup{draft}

\begin{document}

\date{}

\title{\Large \bf \na: Temporal Memory Safety via Robust Points-to Authentication}

\author{
{\rm Reza Mirzazade Farkhani}\\
Northeastern University \\
mirzazadefarkhani.r@northeastern.edu
\and
{\rm Mansour Ahmadi}\\
Northeastern University \\
Mansosec@gmail.com
\and
{\rm Long Lu}\\
Northeastern University\\
l.lu@northeastern.edu
} %

\maketitle
\pagenumbering{gobble}
\setcounter{page}{1}

\begin{abstract}
 
Temporal memory corruptions are commonly exploited software vulnerabilities that
can lead to powerful attacks. Despite significant progress made by decades of
research on mitigation techniques, existing countermeasures fall short due to
either limited coverage or overly high overhead. Furthermore, they require
external mechanisms (e.g., spatial memory safety) to protect their metadata.
Otherwise, their protection can be bypassed or disabled. 

To address these limitations, we present {\em robust points-to authentication},
a novel runtime scheme for detecting all kinds of temporal memory corruptions.
We built a prototype system, called \na, that realizes this scheme on ARM
architectures. 
\na contains a customized compiler for code analysis and instrumentation and a
runtime library for performing the points-to authentication as a protected
program runs. \na leverages the Pointer Authentication Code (PAC) feature, 
provided by the ARMv8.3 and later CPUs, which serves as a simple
hardware-based encryption primitive. \na uses minimal in-memory metadata and
protects its metadata without requiring spatial memory safety. We report our
evaluation of \na in terms of security, robustness and performance using 150
vulnerable programs from Juliet test suite and the SPEC CPU2006 benchmarks. \na
detects all three categories of heap-based temporal memory corruptions,
generates zero false alerts, and slows down program execution by 26\% (this
number was measured based on software-emulated PAC; it is expected to decrease
to 20\% when using hardware-based PAC). We also show that \na incurs
2\% memory overhead thanks to the efficient use of metadata.

\end{abstract}

\section{Introduction}
\label{sec:intro}

Memory corruptions remain to be the most commonly exploited software
vulnerabilities, despite the significant progress made by decades of research on
mitigation techniques. Memory corruptions are caused by 
programming errors (or bugs) that break the type constraints of data in memory. 
They serve as the stepping stone for launching almost
all types of software attacks, from simple stack smashing to heap spray and to
more advanced return-oriented programming (ROP) and code reuse attacks.
Generally, memory corruptions exist in two different forms: {\em spatial} or
{\em temporal}. The former happens when data's spatial boundary is breached. 
The latter is due to data being used out of its life span. 

Temporal memory corruptions may seem less harmful than spatial corruptions.
However, they are being increasingly exploited to bypass the state of the art
defenses against spatial corruptions or control flow manipulations.
Use-after-free (UAF) is the most common type of temporal memory corruption. 
A recent analysis on the Chromium project shows that 50\% of the serious memory safety bugs are UAF issues~\cite{chromium}. To
exploit a UAF, an attacker first plants crafted objects in place of
expired/freed objects and then waits for the vulnerable program to access the
planted objects, and in turn, unknowingly invoke code specified by the attacker.
Double-free and invalid-free are two other types of temporal memory corruptions~\cite{nagarakatte2010cets} that
can provide arbitrary write primitives for attackers.

To counter the powerful and stealthy attacks enabled by temporal memory
corruptions, many mitigation or prevention techniques were proposed recently.
One approach adopted by these
works~\cite{lee2015preventing,younan2015freesentry,van2017dangsan,dang2017oscar,shin2019crcount}
aims to disable dangling pointers, without which UAF and its variants cannot
occur. Though effective, these techniques are either too heavy for real-world
deployment~\cite{lee2015preventing,younan2015freesentry} or limited in their scope of
protection. For instance, DangNull~\cite{lee2015preventing} can only protect
those pointers that reside on the heap. 
Techniques solely focusing on preventing dangling pointers, such as
Oscar~\cite{dang2017oscar}, are unable to prevent invalid-free vulnerabilities.

Another line of works on preventing temporal memory corruptions monitors every
pointer dereference during runtime and ensures that the to-be-dereferenced
pointer indeed points to the expected object (or
type)~\cite{nagarakatte2010cets,burow2018cup,nagarakatte2015everything}. These techniques are, 
in principle, more comprehensive than dangling pointer prevention. However, they
tend to incur heavier runtime overhead. 

Despite the approaches, the aforementioned techniques all require spatial
memory safety to protect their in-memory metadata, whose integrity is critical
for the runtime monitoring. This common requirement underlines two limitations
of these techniques. First, without external protection, they themselves are not
robust against attacks or evasions. Second, requiring spatial safety can
significantly increase the already high runtime overhead. 
Furthermore, many of the existing mitigations against temporal memory corruption,
including~\cite{lee2015preventing,van2017dangsan,younan2015freesentry}, store a
considerable amount of metadata in memory, increasing the memory footprint by as
much as 2 times.

\vspace{1em} 
Motivated by the limitations of previous works (esp. limited coverage,
the requirement of external protection, and high overhead), we present \na, a novel
system for dynamically detecting temporal memory corruptions in user-space
programs. \na follows the approach of runtime dereference checking. Unlike
previous works, \na has built-in protection of its in-memory metadata and thus
obviates the need for external mechanisms to provide spatial memory safety.
Moreover, \na uses a checking scheme that minimizes metadata size and optimizes
metadata placement for better compatibility and handling of data and pointer
propagation.   

Specifically, during the allocation of every heap object, \na assigns a unique
ID to the object and computes a cryptographic authentication code (\ac) based on
the object ID and the base address of the object. \na stores the ID to the
beginning of the object. It stores the \ac to the unused bits of the pointer to
the object. As a result, the pointer is ``tied to'' the object (or the pointee)
at the particular location in memory. This points-to relationship can be verified
during every pointer dereference by re-computing the \ac. An \ac mismatch
indicates a temporal memory safety violation. 

The in-memory metadata of \na include \ac for pointers and IDs for objects.
Obviously, the robustness of the runtime checks hinges on the integrity of the
metadata. \na can detect corrupted or invalid metadata without requiring any
form of spatial memory safety, thanks to the design of \ac. By using a secret
key for computing and verifying \ac, \na prevents attackers from forging or
tampering with metadata. We prototyped \na for the latest ARM architecture and
employ PAC (pointer authentication code)~\cite{pacv83}, a hardware-based
feature, to implement \ac. PAC was originally designed for checking the
integrity of protected pointers and has been enabled on the latest iOS
devices~\cite{a12pac}. We repurposed this hardware feature for performing secure
encryption (i.e., calculating \ac) and storing \ac in unused bits of pointers.  

The in-pointer storage of \ac offers two benefits. First, storing \ac does not
consume additional memory space. Second, an \ac is propagated automatically when
the pointer is copied or moved, without requiring handling or tracking by \na.
An object ID is 8-byte long and is stored at the beginning of the object. This
distributed placement of object IDs, as opposed to centralized storage, makes
the runtime check faster.

In summary, we made the following contributions: 

\begin{itemize}
    \item We designed a novel scheme for dynamically detecting temporal memory
    corruptions, which overcomes the limitations of previous works and achieves
    minimal metadata, full coverage, and built-in security against attacks and
    metadata tempering. 
    \item We built a system for ARM platforms that utilizes PAC to implement the
    detection scheme in an efficient and secure way. 
    \item We evaluated the prototype using standard benchmarks and compared it
    with the state-of-the-art temporal corruption detectors, confirming the
    advantages of our approach. 
\end{itemize}

\section{Background}
\label{sec:background}

\subsection{Exploiting Temporal Memory Bugs}
\label{sec:attacks}

\point{Use-after-free}
If a program reuses a pointer after the corresponding buffer had been freed,
attackers may plant a crafted object in the same memory location, after the free
and before the use, to trick the program into using the crafted object and
consequently perform attacker-specified actions. According to recent reports
\cite{lee2015preventing,caballero2012undangle}, UAF now counts for a majority of
software attacks, especially on browsers, mostly because the deployed attack
mitigations are unable to detect them. Moreover, most of the recent Android
rooting and iOS jailbreaking exploits use UAF as a key part of their attack
flows \cite{lookout}.

\point{Double-free} 
Double-free is a special case of UAF, which occurs when a pointer is freed twice
or more. This leads to undefined behaviors~\cite{cspecification} and can be
exploited to construct arbitrary memory write primitives, with which an attacker
can corrupt sensitive information such as code pointers and execute arbitrary
code.

\point{Invalid-free}
Invalid-free occurs when freeing a pointer that is not pointing to the beginning
of an object or a heap object at all (i.e., freeing a pointer that was not
returned by an allocator)~\cite{nagarakatte2010cets,cspecification}. 
Similar to double-free, invalid-free may allow
attackers to gain arbitrary memory overwrite abilities. The idea of
\textit{House of Spirit}~\cite{houseofspirit} exploitation technique is partly
based on exploiting invalid-free errors.

\subsection{Pointer Authentication Code on ARMv8}
\label{sec:PAC}

Pointer Authentication Code, or PAC, is a new hardware feature available on
ARMv8.3-A and later ARM Cortex-A architectures \cite{armv8a}. PAC is designed
for checking the integrity of critical pointers. Compilers or programmers use
the corresponding PAC instructions to (1) generate signatures for selected
pointers, and (2) verify signatures before signed pointers are used. For
instance, in a typical use case of PAC, compilers insert to programs the PAC
instructions that, during runtime, sign each return address (\ie a special code
pointer) before saving it and check the signature before every function return.
PAC is designed to detect unexpected or malicious overwrites of pointers. It has
been deployed and enabled on the latest iOS devices \cite{a12pac}. 

PAC generates pointer signatures, or authentication codes, using \qa
~\cite{avanzi2017qarma}, a family of lightweight block ciphers. \qa takes two
64-bit inputs (one pointer and one context value), encrypts the inputs with a
128-bit key, and outputs a 64-bit signature. A context value is chosen by the
programmer or compiler for each pointer. A total of five keys can be set by the
OS (\ie code running at EL1) for encrypting/signing different kinds of pointers.
Signatures are truncated and stored in the unused bits of signed pointers (\ie
depending on the virtual address space configuration, 11 to 31 bits in a  64-bit
pointer could be unused). 

Very recently, ARM announced ARMv8.6-A~\cite{armv86}, which introduced some
enhancements to PAC. In ARMv8.3, when a pointer authentication process fails,
the top bits of the invalid pointer is changed to \textit{0x20}, which makes the
pointer invalid to use. In contrast, in ARMv8.6, an exception is thrown when a
pointer authentication fails, which prevents an attacker to brute-force the
correct signature. Another improvement in ARMv8.6 is that a signature is XORed
with the upper bits of the pointer, which help mitigate signature reuse attacks.
At the time of writing this paper, no publicly available hardware or simulator
supports ARMv8.6. Our design and implementation of \na are based
on ARMv8.3. We discuss in \S\ref{sec:instrumt} how our design can be made compatible with ARMv8.6. 

Table~\ref{table:pacinstructions} lists a subset of PAC instructions. Each
instruction serves one purpose (signing or authentication), targets one type of
pointers (code or data), and uses one of the five keys (\ie two keys for each
pointer type plus a generic key). Differentiating pointer types and having
multiple keys help reduce the chance of pointer substitution or reuse attacks. 
The bottom two instructions in
Table~\ref{table:pacinstructions} are special. \pacga is not specific to pointer
authentication and can be used as a data encryption instruction on small data objects (16 bytes at most). It uses the
generic key and outputs a 32-bit cipher to the upper half of a general-purpose
register. {\tt XPAC} removes the signature from a signed pointer without any authentication. Therefore, it does not use any key.
PAC is designed for fast and robust checking of pointer integrity. The signing and
authentication are performed directly by the CPU without any software-level
assistance. The keys are stored in the special CPU registers, which are
accessible only to OS or EL1 code and not visible to user-level code.

\begin{table}[t]
    \centering
    \begin{tabular}{llll} 
    \toprule
    \multicolumn{1}{c}{Instruction}               & \multicolumn{1}{c}{Key Used}             & \multicolumn{1}{c}{Pointer Type}                    & \multicolumn{1}{c}{Purpose}                                 \\ 
    \toprule
    
    \multicolumn{1}{c}{PACIAx} & \multicolumn{1}{c}{Code.A} & \multicolumn{1}{c}{Code} & \multicolumn{1}{c}{Signing}  \\
    \multicolumn{1}{c}{PACIBx} & \multicolumn{1}{c}{Code.B} & \multicolumn{1}{c}{Code} & \multicolumn{1}{c}{Signing}  \\
    \multicolumn{1}{c}{PACDAx } & \multicolumn{1}{c}{Data.A} & \multicolumn{1}{c}{Data} & \multicolumn{1}{c}{Signing}                      \\ 
    \multicolumn{1}{c}{PACDBx } & \multicolumn{1}{c}{Data.B} & \multicolumn{1}{c}{Data} & \multicolumn{1}{c}{Signing}                      \\ 
    \hline
    
    \multicolumn{1}{c}{AUTIAx } & \multicolumn{1}{c}{Code.A} & \multicolumn{1}{c}{Code} & \multicolumn{1}{c}{Authentication}  \\
    \multicolumn{1}{c}{AUTIBx } & \multicolumn{1}{c}{Code.B} & \multicolumn{1}{c}{Code} & \multicolumn{1}{c}{Authentication}  \\
    \multicolumn{1}{c}{AUTDAx  } & \multicolumn{1}{c}{Data.A} & \multicolumn{1}{c}{Data} & \multicolumn{1}{c}{Authentication}                      \\ 
    \multicolumn{1}{c}{AUTDBx  } & \multicolumn{1}{c}{Data.B} & \multicolumn{1}{c}{Data} & \multicolumn{1}{c}{Authentication}                      \\ 
    \hline
    
    \multicolumn{1}{c}{PACGA} & \multicolumn{1}{c}{Generic} & \multicolumn{1}{c}{Generic} & \multicolumn{1}{c}{General}  \\
    \hline
    
    \multicolumn{1}{c}{XPAC} & \multicolumn{1}{c}{-} & \multicolumn{1}{c}{-} & \multicolumn{1}{c}{Sig. stripping}  \\
    \hline
    
    \end{tabular}
     \caption{PAC-related instructions.\\ }
     \label{table:pacinstructions}
\end{table}

PAC was originally designed for checking the integrity of pointers and has been mostly used for protecting code pointers. 
We use PAC as a simple hardware-based primitive for efficiently and securely
computing \ac. The \ac is computed and verified based on our novel scheme designed for
detecting temporal memory corruptions. 
Compared with PAC, \na achieves a security goal (\ie enforcing temporal memory safety) that is orthogonal to, and
broader than, the original purpose of PAC (\ie checking pointer value integrity).
The recent UAF vulnerability in iOS (CVE-2019-8605~\cite{SockPuppet}) 
is a real example that shows PAC is unable to
prevent temporal memory corruptions, which are exploited for jailbreaking or compromising iOS devices. 
In contrast, \na is designed to stop temporal memory corruptions, which remain a type of
commonly exploited vulnerabilities today.

\subsection{Fixed Virtual Platforms (FVP)}
\label{sec:fvp}
The ARMv{8.3}-A architecture (including PAC) was announced in late 2016 and is
expected to enter mass production in 2020 to replace the current
mainstream mobile architecture, namely ARMv{8.0}-A. At the time of writing,
no development boards or commercially available SoC (Systems-on-Chip) use
ARMv{8.3}-A. Apple's latest iOS devices, using the A12
Bionic SoC, is based on ARMv{8.3}-A and supports PAC. However, the SoC
and OS are proprietary and cannot be used for testing the prototype of \na.  

ARM offers so-called Fixed Virtual Platforms (FVP) for to-be-released
architectures~\cite{fvp}. FVP is a full-system simulator that includes
processors, memory, and peripherals. It is a functionally accurate model of the
simulated hardware. FVP allows for the development and testing of drivers,
software, and firmware prior to hardware availability. It is widely used in the
industry. 

Following this standard practice, we used the ARMv{8.3}-A FVP when building and
evaluating our prototype system. Thanks to FVP's functional accuracy, the
evaluation results obtained on FVP are expected to be close to those obtained on
actual hardware. We discuss more the implementation and evaluation in
\S\ref{sec:impl} and \S\ref{sec:eval}, respectively.

\section{Design}
\label{sec:design}

\subsection{System Overview} 
The goal of \na is to dynamically detect temporal memory corruptions in the
heap. The high-level idea is that, upon each pointer dereference (or
pointer-based object access), temporal memory corruption can be detected by
checking (1) whether the pointer is pointing to the original or intended object,
and (2) whether the metadata or evidence proving the points-to relationship
is genuine. 

\begin{figure*}[!t]
    \centering
    \includegraphics[scale = 0.8]{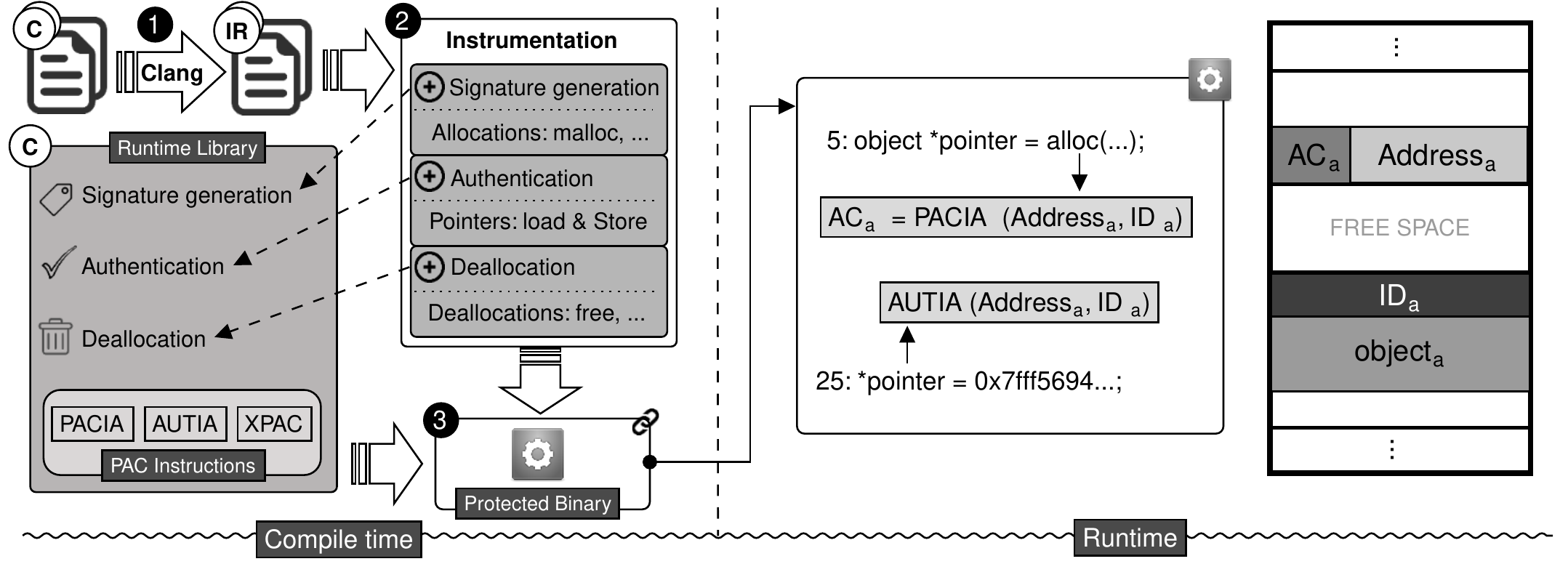}

    \caption{During the compile-time (left), \na instruments a {\tt C} program. It places the hooks in the program that are needed for the \na runtime library to detect heap-based violations of temporal memory safety. The \na library uses a novel authentication scheme (\S\ref{sec:scheme}) that verifies the points-to relationship from pointers to pointees. During runtime (right), the \na library generates signatures (\ac) for heap objects and their pointers upon memory allocations, checks signatures upon pointer dereferences, and invalidates signatures upon object deallocations (\S\ref{sec:instrumt}). The scheme uses PAC as a simple building block for hardware-based metadata signing, storing (for pointers only), and verification. }

    \label{fig:systemoverview}
\end{figure*}

Although the high-level idea is conceptually straightforward, how to realize it
in an efficient and secure way is in fact challenging. What metadata are needed
for establishing the points-to relationship? How are they computed and where are
they stored? How can their integrity be verified? Answers to these design
questions determine the efficiency and robustness of \na. For instance,
recording too much metadata leads to unnecessarily big memory footprint and
redundant checks. Storing metadata separately from objects and pointers may ease
metadata protection but significantly increase the overhead for locating and
accessing metadata. Storing metadata in-place allows for fast access, but pointer
arithmetics may complicate locating metadata. Moreover, in-place metadata is
hard to protect and can be easily corrupted.

Our points-to authentication scheme overcomes these challenges and the limitations
of previous works. \na randomly generates an ID for each heap object upon its
allocation. It also computes a cryptographic authentication code (\ac) based on
the object ID and the base address of the object. The object ID, stored at the
beginning of the object, and the \ac, stored in the unused bits of the object's
pointer, together serve as the metadata to establish the verifiable points-to
relationship between the object and its pointers. Furthermore, \na can detect
forged or corrupted metadata as long as the key for computing \ac remains
confidential and the \ac computation can only be performed by \na. We discuss
the detailed design of the points-to authentication scheme in \S\ref{sec:scheme}.

Our implementation of the points-to authentication scheme takes advantage of the
PAC feature on ARM architectures. \na uses PAC as a simple primitive, provided
by hardware, for computing and checking \ac and securing the key. We discuss in
\S\ref{sec:instrumt} the use of the PAC instructions and the compiler-based code
instrumentation.

\begin{figure*}
    \centering
    \captionsetup[subfigure]{justification=centering}
    \begin{subfigure}[b]{0.31\textwidth}
        \begin{lstlisting}
int* ptr = (int*)malloc(10);
int* qtr = ptr;
...
if (error) {
    free(ptr);
    ptr = null;
}
...

if (log)
    logError("Error", qtr);
        \end{lstlisting} 
        \caption{Use-After-Free}
        \label{fig:uaf}
    \end{subfigure}
    \hspace{0.005\textwidth}    
    \begin{subfigure}[b]{0.31\textwidth}
        \begin{lstlisting}
int* ptr = (int*)malloc(10);
int* qtr = ptr;
...
if (error) {
    free(ptr);
    ptr = null;
}
...

cleanCache:
    free(qtr);            
\end{lstlisting} 
        \caption{Double-free}
        \label{fig:doublefree}
    \end{subfigure}
    \hspace{0.005\textwidth}
    \begin{subfigure}[b]{0.32\textwidth}
       \begin{lstlisting}
char* ptr = (char*)malloc(10);
for (; *ptr != '\0'; ptr++){
    if (*ptr == SEARCH_CHAR)
    {
        printf("Match!");
        break;
    }
}
...
cleanCache:
    free(ptr)
\end{lstlisting} 
        \caption{Invalid-Free}
        \label{fig:invalidfree}
    \end{subfigure}
    \caption{Examples of double-free, use-after-free and invalid-free 
    temporal memory corruptions, which are undetectable by pointer 
    integrity approaches but detectable by \na.}
    \label{fig:exmplevul}
\end{figure*}

Figure~\ref{fig:systemoverview} presents an overview of the \na system. The \na
compiler instruments a protected application by inserting a runtime library and
placing necessary hooks before selected \texttt{load} and \texttt{store}
operations. During runtime, the \na library checks the instrumented,
pointer-based load/store operations. The checking is based on a novel scheme
that verifies the points-to relationship and the metadata integrity. It uses PAC
as the hardware-based authentication primitive. \na also installs a tiny OS
patch for managing PAC encryption keys, which are only accessible from the
kernel-space (or EL1) as enforced by the architecture. We discuss the design
details after explaining the threat model.

\subsection{Threat Model}
\label{sec:threatmodel}

We adopt a threat model common to user-space dynamic memory error checkers. We
trust the OS and the underlying hardware (\ie the TCB). It is technically
possible to reduce or remove the trust on OS if a more privileged entity can
protect the PAC key management routine (\eg a hypervisor or EL2), which however
is out of the scope for our current design. Our threat model also assumes that
the basic defenses against code injection and modification are in place (\eg DEP
and read-only code). This assumption is realistic because such defenses are
universally enabled on modern OSes. They are needed for protecting code
instrumented by \na (\eg inline checks cannot be removed or uninstrumented code
cannot be injected). 

Our threat model also assumes that attackers cannot perform arbitrary memory
read when exploiting temporal memory errors.  
Arbitrary memory read would allow an attacker to read a legitimate \ac in memory and
possibly reuse it, thus bypassing the security check.
Assuming the absence of arbitrary memory read in our context is acceptable
because finding an \ac as well as its corresponding object ID in memory can be
quite challenging due to ASLR and the indistinguishability between \ac or ID
values and other in-memory data. Moreover, based on previous
research~\cite{di2015elf,bittau2014hacking} and real-world
attacks~\cite{fireeye}, attackers often exploit temporal memory errors as a stepping stone to obtain arbitrary memory read abilities, as shown in the high-profile
WhatsApp double-free (CVE-2019-11932), Internet Explorer use-after-free
(CVE-2013-3893) and iOS use-after-free (CVE-2019-8605 ~\cite{SockPuppet,surveyios}) exploits.
Therefore, it is realistic to assume attackers cannot perform arbitrary memory
read while exploiting temporal memory errors. 

Previous works on temporal memory safety
\cite{lee2015preventing,van2017dangsan,nagarakatte2010cets,shin2019crcount,younan2015freesentry}
made all the above assumptions as we do. Additionally, they assumed the absence
of spatial memory violations or required an external spatial safety mechanism to
protect their metadata. In contrast, \na does not make this assumption or
require external spatial safety enforcement. We relaxed the threat model used in
the previous work by allowing arbitrary memory overwrite, which an attacker may
use to corrupt the metadata. Unlike the previous work, \na has built-in metadata
integrity check and is therefore robust against metadata corruption caused by
spatial memory errors or attacks.

One might argue that \na can be bypassed by attackers who are able to perform
arbitrary memory read and write at the same time. While this argument is
technically true, such a powerful attacker does not need to exploit temporal
memory vulnerabilities at all, or try to bypass \na, because she already has the
ability to directly mount the final-stage attacks, such as code injection or
data manipulation.

\subsection{Example Vulnerabilities}

Before describing our points-to authentication scheme, we present three simple
examples of temporal memory corruption below, which help explain why PAC can
reliably detect them. 

\point{Use-after-free vulnerability} Figure \ref{fig:exmplevul} (a) is a typical
example of UAF, where a pointer is used after its pointee has been freed.  In
this case, \texttt{qtr}, an alias of \texttt{ptr},  is used at Line 11 after
\texttt{ptr} has been freed at Line 5. Although the programmer nullified the
\texttt{ptr} at line 6, due to the aliasing, UAF still exists.

\point{Double-free vulnerability} Figure \ref{fig:exmplevul} (b) shows a code
snippet where a pointer can be freed twice, which may lead to undefined
behaviors, including arbitrary memory writes.

\point{Invalid-free vulnerability} 
Figure \ref{fig:exmplevul} (c) demonstrates a case where a pointer is freed
while it is not pointing to the beginning of a buffer. This is a special type of
temporal memory corruption~\cite{nagarakatte2010cets,cspecification}. 

\begin{figure}[!t]
    \center
    \includegraphics[scale = 0.75]{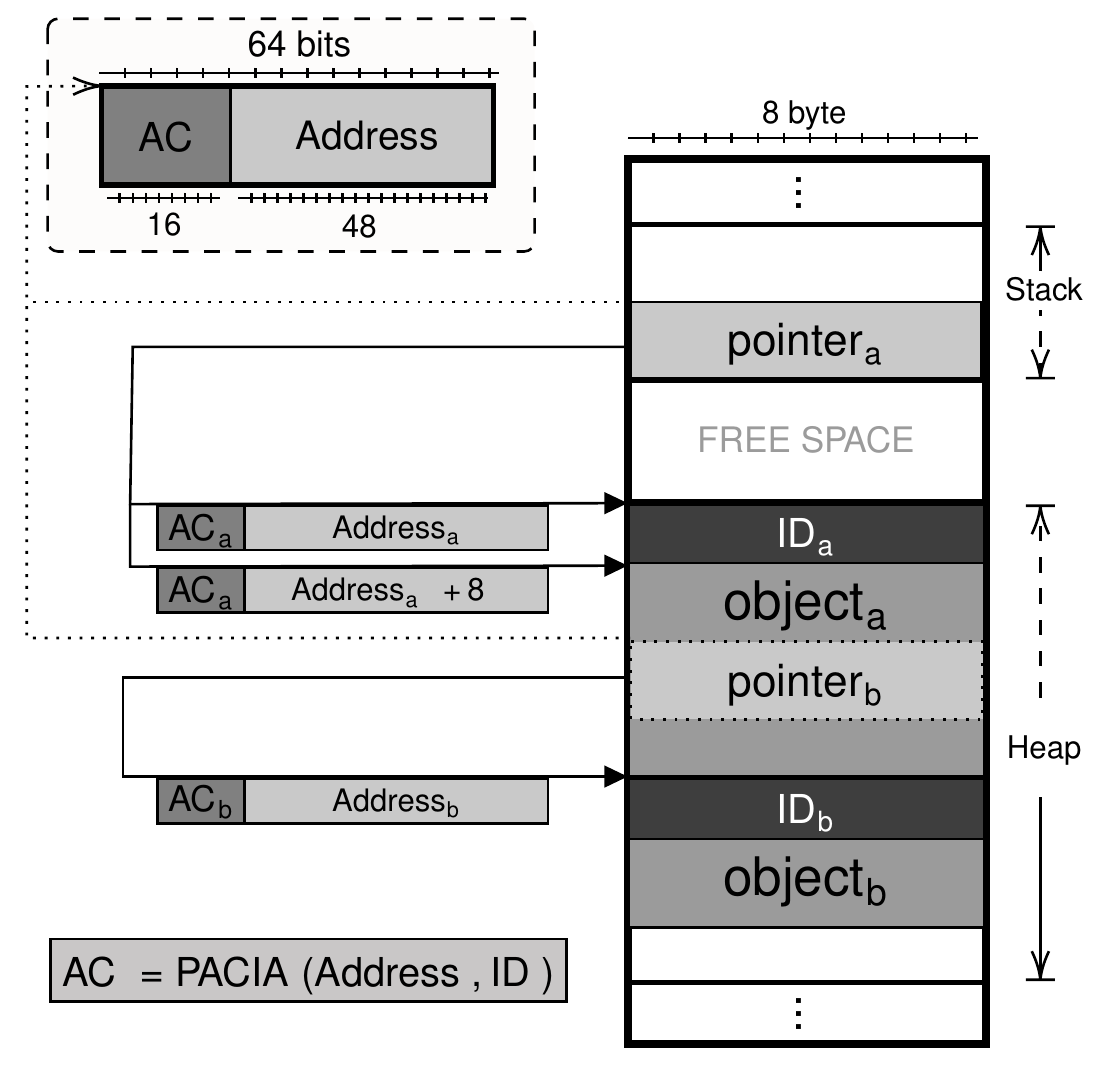}
    \caption{Authentication code (\ac) and object metadata (ID) defined 
    by \na for pointers and objects. The object metadata is stored
    in the 8-byte memory proceeding the object. The \ac is stored 
    in the unused bits of the pointer, which is 16-bit long. }
    \label{fig:unifieddesign}
\end{figure}

\subsection{Points-to Authentication Scheme}
\label{sec:scheme}

Our authentication scheme applies to two types of data: objects and data
pointers. {\em Objects} are dynamically allocated data on the heap. 
{\em Data pointers} reference the addresses of objects (we only consider
pointers to heap objects in this paper). \na verifies the identity of every
object and the points-to relationship before it is accessed through a pointer.
This verification relies on the \ac (or authentication code) generated for the
object and stored in its pointers.

The {\em ID} of an object is saved as inline metadata immediately before the
object in memory (Figure \ref{fig:unifieddesign}). The ID establishes unique
identities for objects and allows for binding pointers to their referenced
objects (\ie making the points-to relationship verifiable), which is essential
for detecting temporal memory corruptions. Figure \ref{fig:unifieddesign} (lower
right) shows two objects in the heap with their metadata. The ID is a 64-bit
random value generated at the allocation of the object. An \ac is 16-bit long
and stored in the unused bits of a pointer (i.e., 48 effective bits in a
pointer). Unlike the previous works such as DangNull~\cite{lee2015preventing},
which only protect pointers residing in the heap, \na authenticates data pointers
stored everywhere in memory, including heaps, stacks and global regions.

Next, we explain the definition and calculation of \ac.  We then discuss in
\S\ref{sec:instrumt} the runtime \ac generation and the checking mechanism.

\point{Data Pointers}
\ac essentially binds a data pointer to its pointee and makes the binding
verifiable. \ac encodes: (1) the identity of the pointee object, and (2) the
base address of the pointee. The ID and the base address together uniquely
identify an object in time and space. This definition not only makes the
points-to relationship easily verifiable, but also mitigates metadata reuse
attacks. Figure~\ref{fig:unifieddesign} (bottom left) shows the computation of
\ac using the {\tt PACIA} instruction. \na performs this computation when an object is
allocated. When an object is deallocated or reaches the end of its life cycle,
\na simply invalidates its ID (setting it to zero). Upon each pointer
dereference, \na recomputes the \ac and compares it with the \ac stored in the
pointer. A mismatch indicates a temporal memory safety violation. No temporal
memory corruption can happen without failing the points-to authentication.

In our scheme, less memory is used for storing the metadata for both pointers
and objects than most previous works. Furthermore, there is no assumption that
the metadata cannot be tampered with. Last but not least, \na can find the base
address of an object reliably with the help of PAC. This is necessary for
supporting pointer arithmetic operations, which may shift a pointer to the
middle of its pointee, than thus, fail a naive authentication that simply takes
the pointer value as the object base address. 
We discuss the details in \S\ref{sec:instrumt}.

The PAC instruction encrypts/signs the inputs (i.e., object ID and base address)
using a data pointer keys ({\tt data.A} or {\tt data.B}) and saves the truncated
ciphertext to the unused bits in the pointer. Therefore, unlike an object, \na
does not need to use extra space for storing \ac for pointers. An \ac is
generated whenever a pointer takes a new value, which can happen at object
allocation or when the pointer is re-assigned to another object (\eg via the
reference operator ``{\tt \&}'').

After a pointer becomes stale when its pointee is freed, any dereference of the
dangling pointer will trigger an object ID mismatch, due to either the
invalidated ID of the freed object, or a different ID of a new object allocated
at the same location. Other temporal memory errors, such as double-free and
invalid-free can be detected by \na in the same way.

\point{Code Pointers}
Checking the integrity of code pointers is an intended use of PAC and is fairly
straightforward. Unlike data pointers, we do not define our own \ac for code
pointers. \na is fully compatible with the intended use of PAC for code pointers
for preventing control flow hijacking attacks. They can be used together to
thwart a broad range of attacks. We do not consider or claim code pointer
authentication as a contribution to this work. For the rest of the paper, we
focus our discussion on authenticating the points-to relationships while
referring readers to the PAC documentation \cite{pac,pacv83} and PARTS
\cite{liljestrand2018pac} for code pointer authentication.

\subsection{Compiler-based Code Instrumentation \& Runtime \ac Checking}
\label{sec:instrumt}

To apply the points-to authentication scheme to a given program, \na takes the
general approach of inline reference monitoring. Via a custom compilation pass
added to LLVM \cite{llvm}, \na instruments the program so that
\ac can be generated and checked at the right moments
during program execution. The instrumentation is performed at the LLVM bitcode
level, which is close to assembly code while retaining enough type and semantic
information for our code analysis and instrumentation.
The instrumentation sites are carefully selected to minimize the interception of
program execution. Below we discuss in detail the code instrumentation needed for each
type of operation on \ac.

\point{AC Generation}
During runtime, \na needs to generate \ac for data pointers whenever a new 
points-to relationship is created. To this end, during
compilation, \na performs two types of instrumentation. First, it instruments
all essential API for heap memory allocation, including {\tt malloc} (the
dynamic allocator for heap objects), {\tt calloc} and {\tt realloc}. \na only
works on user-space programs and we assume the {\tt ptmalloc} allocator is used.
This instrumentation allows \na to intercept all memory allocations, where the
object ID is generated and the \ac for the pointer is computed as follows:

\lstset{
language = C,
numbers=left,
xleftmargin=2em,
frame=single,
framexleftmargin=2em,
escapeinside={<@}{@>},
}

\begin{lstlisting}
/* Computing AC for Data Pointer */
ID = <@\textcolor{purple}{\textbf{RandomID}}@>() // 64-bit
AC = <@\textbf{PACIA|B}@> <BasePointer><ID>
\end{lstlisting}

Second, \na instruments object deallocation sites, like {\tt free} (heap object
deallocation). At an object deallocation site, \na simply sets the object ID to
zero, which invalidates the object and thus prevents any further use of the object.
Figure~\ref{fig:unifieddesign} (upper right) shows an example pointer and its
\ac. The base address and the ID of the
pointee are used as the two inputs to the {\tt
PACIA|B} instruction to generate the \ac:

\point{AC Checking}
\na performs points-to authentication by checking the \ac whenever a
pointer-based data access happens (or a pointer reference occurs). During
compilation, \na  instruments LLVM {\tt load} and {\tt GetElementPtr}
instructions for pointers. For simplicity, we generally refer to both as load in
our discussion.
\na verifies the integrity of the pointer and 
authenticates the \ac of the pointer 
value as follows:

\begin{lstlisting}
/* Authenticating AC for Data Pointers */
ID = <@\textcolor{purple}{\textbf{getID}}@>(Pointee) // Pointee is an object
<@\textbf{AUTIA|B}@> <BasePointer><ID>
\end{lstlisting}

Due to pointer arithmetics, a (legitimate) pointer may sometimes point to the
middle, instead of the base, of its pointee. Therefore, during \ac checking, \na
cannot simply use the value of the pointer as the base address of the
to-be-accessed object. A naive solution to this problem is to use additional
metadata for recording the object base address for each pointer. However, this
not only increases space overhead but creates a more challenging problem of
propagating the metadata as pointers are copied or moved.   

\point{Backward Search}
\na finds the base address of an object during runtime without requiring any
additional metadata. For each \ac checking, \na, by default, uses the pointer
value as the object base address. If the check fails, two possibilities arise.
First, the pointer is valid but is pointing to the middle of its pointee (i.e.,
its value is not the base address, hence the mismatched \ac). Second, the
pointer is invalid and a temporal memory violation is about to happen. When
encountering a failed \ac check, \na initially assumes that the first possibility
happened. It then starts a backward search from the current pointer location for
the based of the object. Since objects are 16-byte aligned in memory,
the backward search only looks for the object base addresses divisible by 16. 
This
optimization makes the backward search fast. The search terminates when (1) an \ac match occurs (\ie the
correct Object ID and the base address are found), or (2) the search has
exceeded the max distance or reached invalid memory, in which case a true
temporal memory error is detected. 

Our backward search scheme is tested, and works well, on ARMv8.3-A (the latest
Cortex-A architecture available today). It is worth noting that the PAC
instructions in the future ARMv8.6-A architecture may generate an exception when
an authentication fails \cite{armv86}. Our backward search scheme can work with
the exception-enabled PAC instructions by having a tiny kernel patch that
masks/disables the corresponding exceptions~\cite{exceptionmask} during the
(transient) backward search window. The exception masking code is only callable
within the backward search function and thus cannot be abused by attackers. We
are unable to evaluate this patch due to the lack of hardware or
simulator for ARMv8.6. The rest of the \na design and implementation is
compatible with ARMv8.6.

\begin{table*}[ht]
\resizebox{1\textwidth}{!}{  
\centering
\begin{tabular}{l|ccccc}
\hline\toprule
       & CETS \cite{nagarakatte2010cets} & DangNull \cite{lee2015preventing} & DangSan \cite{van2017dangsan} & CRCount \cite{shin2019crcount}  & \na   \\
\hline
Allocation    &  Generate lock \& key   & Register pointer & Register pointer  & Generate reference counter  & Generate ID \& AC       \\
Pointer dereference: *p    &   Comparison of key and lock value  & No check & No check & No check & Points-to authentication         \\
Copy ptr arithmetic:  p = q+1  & Propagate lock address and key   & Update register ptr & Update register ptr & Update reference counter  &  No cost   \\
Deallocation     &   Invalidate lock  & Invalidate pointers & Invalidate pointers & Delayed deallocation & Invalidate ID        \\
Memory overhead    &  O (\# pointers) & O (\# pointers) & O (\# pointers) & O (\# pointers) + Mem leaks & O (\# objects)     \\
Metadata handling    &  Disjoint & Disjoint & Disjoint & Disjoint & Inline     \\
Metadata safety guarantee   &  \ding{53} & \ding{53} & \ding{53} & \ding{53} &   \ding{52}   \\

\bottomrule\hline
\end{tabular}
} \caption{Comparison of our approach with the closely related works, in
terms of the use/check, management, and protection of the metadata. 
}
\label{tab:comparison}
\end{table*}

\point{Metadata propagation}
Thanks to our in-pointer storage of \ac, when a pointer is copied or moved, the
metadata of the pointer is automatically propagated without any special handing
by \na or any software. As for metadata for objects (\ie IDs), they are not
stored inside objects and thus are not automatically propagated during object
duplication or movement. However, this is intended---object metadata should not
be propagated when objects are copied or moved. This is because in our points-to
authentication scheme, an object ID is assigned to and associated with the
allocated buffer, rather than the data stored in that buffer. In contrast,
previous works on temporal memory error detection, such as CETS
\cite{nagarakatte2010cets}, require special handling of metadata propagation at
the cost of degraded runtime performance and limited data compatibility.

\point{Handling deallocation}
In contrast to pointer dereferencing, where a pointer can point to the middle of
an object, for the deallocation procedure, the pointer should always point to
the beginning of the object. Otherwise, invalid-free occurs, leading to
undefined behaviors and temporal memory errors~\cite{cspecification}. Based on
this fact, \na only performs one round of \ac checking without the backward base
address search. If the authentication fails at a deallocation site, it is either
a double-free or an invalid-free error. If the authentication succeeds, \na
simply sets the object ID to zero (\ie invalidation) and lets the program
execution continue.

\point{Handling reallocation}
During reallocation, the base address of an object may or may not change
depending on the size of the object and the layout of the memory. \na handles
reallocation by instrumenting {\tt realloc}. If the base of an object has changed,
\na nullifies the ID of the old object, generates a new ID, and computes a new \ac
for the new base pointer. As a result, the existing (stale) pointers
to the old object become invalid and cannot be used anymore.

\point{External/uninstrumented Code}
During compilation time, \na treats as a blackbox externally linked code or code
that cannot be instrumented. This design enables backward and external
compatibility. 
\na instruments the entries to such blackboxes so that immediately before an
object or pointer flows into a blackbox (\eg as an argument to an external
function call such as {\tt memcpy}), \na authenticates the pointer and then
strips off its \ac, which can be done efficiently using the {\tt XPAC}
instruction. Conversely, when a pointer returns from a blackbox,
\na generates the \ac for it, whose subsequent uses are subject to checks.

\subsection{Optimizations}
\label{sec:optmization}
\point{Unnecessary Checks} 
We optimize the instrumentation strategy by avoiding insertions of unnecessary
checks during compilation. The optimization is inspired by the fact that, for
any valid pointer, UAF and other temporal memory violations cannot happen
through the pointer until it is being freed or later. Therefore, it is not
necessary to perform points-to authentication on any use of a pointer that can
only take place before the pointer is freed. Obviously, detecting all such
pointer uses in a program is an untractable
problem~\cite{Landindecidability,Ramalingamaliasing}, which requires perfect
alias analysis. However, we can solve this problem within the scope of a
function by performing conservative intra-procedural analysis. By tracking a
pointer's def-use chain inside a function, we can identify a set of use sites
where the pointer and its aliases have not been free or propagated out of the
function. \na can safely ignore these use sites during instrumentation (\ie no
runtime check is needed).

\begin{figure}[ht]
    \begin{lstlisting}
void quantum_gate2 (quantum_reg *reg){    
int i, j, k, iset;
int addsize=0, decsize=0;

if(reg->num > reg->max )
      printf("maximum",reg->num);

else {
  for(i=0; i<(1 << reg->hashw); i++)
    reg->hash[i] = 0;
  
  for(i=0; i<reg->size; i++)
    quantum_add_hash(reg->node[i].state, i, reg);
    ...

    \end{lstlisting}
    \vspace{-0.1em}
    \caption{Optimization in \na. This example shows that the {\tt reg} 
    pointer is used multiple times in this function. Since the 
    pointer is authenticated before passing to the {\tt quantum\_gate2} function, 
    no check on it is needed until Line 13 where the pointer is passed 
    to another function as an argument. 
    Due to the limitation of intra-procedural analysis, we cannot track the pointer
    into the {\tt quantum\_add\_hash} function to make sure that it is not being freed.
    Therefore, After this point, all the temporal checks will be in place.}

    \label{fig:optimization}
   \vspace{-0.1em}
\end{figure}

Figure~\ref{fig:optimization} demonstrates an example of how redundant checks
are removed by optimization. In this example, all checks on \texttt{reg} up
to Line 13 are unnecessary and are omitted by \na.
Note that this optimization only works on single-threaded programs.
We also extend this optimization to the implementation level. Some frequently
used \textit{glibc} functions such as {\tt printf} and {\tt strcpy} never free
pointers passed to them as parameters. Therefore, we whitelist such functions
and allow the intra-procedural discovery of safe pointer uses to continue beyond
such functions.

\point{Global objects} Performing temporal checks on pointers to global objects is also unnecessary
because such objects are never deallocated. \na detects those address taken
global objects that can be determined statically during the compile-time and
remove the checks for them.

\subsection{Design Comparison}
In Table~\ref{tab:comparison}, we compare \na  with closely related works in
terms of the use/check, management, and protection of the metadata. \na uses
inline metadata, which makes the access fast because no heavy lookup is needed.
Thanks to the inline metadata, the memory overhead of \na is low and there is no
complex handling needed for pointer arithmetics and metadata propagation. \na
uses PAC to compute and secure metadata without requiring external spatial
safety schemes.

\section{Security Analysis}
\label{sec:sec}

An attacker may attempt to evade \na with the goal of causing temporal memory corruption
without being detected. We analyze the possible attacks permitted by our threat
model and explain how the design of \na prevents them. Since \na performs
load-time authentication and our threat model assumes attackers capable of
arbitrarily writing to data memory (\eg by exploiting certain vulnerabilities),
the attacker essentially needs to somehow generate the correct \ac for the data pointer
that she writes before the data is used by the target program or checked by \na.
We note that code inject or modification is not allowed under our threat model
thanks to DEP and the read-only code region. We identify the following ways
that attackers may try to forge the \ac. 

\point{Directly generating AC}
One intuitive evasion of \na is to generate the \ac for the attacker-supplied
data, either offline or dynamically. Offline \ac generation does not work
because the set of keys used for calculating \ac is dynamically generated for
each program execution or process and is not static. Alternatively, the attacker
may try to directly generate \ac on the fly while the target program is running.
This is impossible either because the PAC keys are stored in the special CPU
register and not accessible from the user space, even if the attacker has the
arbitrary memory read capability. Moreover, the attacker cannot inject code and
thus cannot directly calculate \ac using injected PAC instructions. 
Also, brute-force is not applicable in this context because one wrong guess can lead to a crash of the process.

\point{Reuse PAC instructions}
The attacker's next possible move could be to reuse the existing PAC
instructions already loaded in the memory (\eg those used by \na) for
calculating \ac on injected data. However, our system can easily get merged with
the standard use of PAC for protecting code pointers as well. Therefore, code reuse
attacks are prevented thanks to the code pointer integrity check by PAC (\ie any
corrupted return addresses or call/jump targets trigger authentication failures
and are detected before the program control flow is hijacked).

\point{ID spray}
Another possible attack vector is spraying the ID into the object to misguide
the dangling pointers that are pointing to the middle of object. The design of
\na considers this attack. Since the \ac is bound to the beginning address of
an object, even if the correct ID is found in the middle of object,
the authentication will fail.

\section{System Implementation}
\label{sec:impl}

We built a prototype for the \na system, including (\rom{1}) a customized
compiler for instrumenting and building \na -enabled programs, (\rom{2}) a
runtime library, linked to instrumented programs, for performing dynamic \ac
generation and authentication, and (\rom{3}) a set of bootloader and Linux
patches necessary for configuring the CPU and enabling the PAC
feature~\cite{bootwrapper,patchkernel}. All the system components are
implemented in C/C++ with a small set of inline assemblies that directly use the
PAC instructions. The \na LLVM pass is approximately 2K lines of C++ code and
the runtime library is 1K lines of C code. The current implementation supports C
programs. It is based on {\tt ptmalloc} memory allocator from \textit{glibc}.
It supports all common memory allocation APIs, such as {\tt malloc}, {\tt calloc},
{\tt realloc} and {\tt free}.

\point{Customized Compiler}
\label{sec:compiler}
Our compiler is based on LLVM 6.0, which already has basic assembler and
disassembler support for PAC on ARMv8.3-A. We built the code analysis and
instrumentation logic (\S\ref{sec:instrumt}) into an LLVM transform pass. It
operates on the LLVM bitcode IR. At each instrumentation site, such as pointer {\tt
load} and {\tt store}, it inserts a call, based on the type of the instrumented
instruction, to the \na runtime library.

\point{Runtime Library}
The runtime \ac checking logic is built into a dynamically linkable library. It
exposes the call gates for the instrumented code to invoke the \ac generation and
authentication routines. These routines calculate or check \ac for different
scenarios, as describe in \S\ref{sec:scheme} and \S\ref{sec:instrumt}. The
library does not maintain any data internally thanks to the in-place storage of
\ac and the OS-managed PAC keys. Therefore, no data inside the library needs to
be protected or verified. However, we do enable code pointer integrity checking using
PAC when compiling the library, which ensures that no control flow hijacking can
occur while the library code is running.

\point{OS and bootloader patches}
By default, PAC instructions (except for {\tt PACGA} and {\tt XPAC}) are
disabled. According to the ARMv8 reference manual \cite{armv8manual}, to use all
PAC instructions and the corresponding key slots, the OS needs to set to 1 the
{\tt EnIA, EnIB, EnDA, EnDB} fields in the {\tt SCTLR\_EL1} register.
Additionally, the {\tt SCR\_EL3.APK} and {\tt SCR\_EL3.API} registers need to be
set to 1 during the system booting stage. These configurations are necessary to
fully enable the PAC hardware extension. The OS also needs to generate and
manage PAC keys for each process (only OS or code running at EL1 is allowed to
manage PAC keys). We implemented these configurations and tasks via two small
patches to the bootloader \cite{bootwrapper} and the Linux kernel. These small
patches do not interfere with any bootloader or OS functionalities because
(\rom{1}) the configured register fields are reserved exclusively for PAC, and
(\rom{2}) the added PAC key management routine does not interact with the rest
of the OS.

We built the patched bootloader and kernel into a system image,
which was then installed on the ARMv8.3-A FVP. As discussed in \S\ref{sec:fvp},
FVP is the functional-accurate whole-system simulator for ARM architectures,
which emulates processors, memory, and peripherals.  We used this prototype and
environment for evaluating \na. 

\section{Evaluation}
\label{sec:eval}

In this section, we evaluate the prototype of \na in terms of security, runtime
overhead and memory overhead. The security evaluation (\S\ref{sec:evalsec}) was conducted on the ARM FVP
simulator. The performance evaluation (\S\ref{sec:evalperf}) was performed on a Raspberry Pi 4 with
ARMv8-A Cortex A53 processor (1.5GHz) and 4GB memory, running Gentoo 64-bit
Linux (v4.19). We explain the rationale behind this setup in \S\ref{sec:expsetup}.

Our experiments aim to show: (\rom{1}) whether
\na detects temporal memory corruptions such use-after-free, double-free and invalid-free;
(\rom{2}) how much performance overhead \na incurs during runtime; (\rom{3}) how
much memory overhead \na incurs during runtime. We used Juliet test suite
~\cite{black2018juliet} and four real CVEs for security experiments. To evaluate
the runtime and space overhead, we used SPEC CPU2006.

\subsection{Experiment Setup and Methodology}
\label{sec:expsetup}

We performed the security evaluation (\S\ref{sec:evalsec}) on the FVP simulator
that supports PAC. At the time of writing, no publicly available development
board supports ARMv8.3 or PAC instructions. Although Apple's A12 Bionic SoC
supports PAC instructions, it is a proprietary implementation and we were not
able to instrument and run the benchmarks on top of that. 
We patched the bootloader and OS in the FVP image as described in \S\ref{sec:impl}.

We conducted the performance evaluation (\S\ref{sec:evalperf})  on a Raspberry
Pi 4 (ARMv8-A Cortex A53), rather than FVP. This change of platform is necessary because
the benchmarks (SPEC CPU2006) are too heavy to run on the
FVP---they often crash or halt the simulator. To allow \na to run on the Raspberry
Pi, which does not support PAC, we implemented in software the three PAC
instructions used by \na, namely \texttt{PACIA}, \texttt{AUTIA}, and
\texttt{XPAC}. The input/output syntax of these functions is identical to that
of the original PAC instructions. Figure~\ref{fig:paciaimplementation}
demonstrates the implementation of \texttt{PACIA} instruction as a C function.
The other PAC instructions are implemented based on \texttt{PACIA}. 
It is worth noting that our software PAC implementation does not contain the
exact cryptographic algorithm (\textit{QARMA}) used in PAC instructions. This is
because a software implementation of \textit{QARMA} would be much slower than
the hardware implementation and thus make it difficult to measure the real
performance overhead caused by \na. Instead, we chose a simple encryption
and \ac computation, keeping the overhead comparable to hardware-based
encryption and allowing the performance evaluation to focus on the overhead of
\na itself. 

\begin{figure}
    \begin{lstlisting}
long MASKBITS = 0b000...000111111111111111;
void* __pacia(void* ptr,int id){
    long ptrbits = (unsigned long)ptr & MASKBITS;
    long idbits = id & MASKBITS;
    long signature = ptrbits ^ idbits;
    signature = signature << 48;
    unsigned long ptrWithSign = (unsigned long)ptr | signature;
    return (void*)ptrWithSign;
}
    \end{lstlisting}
    \vspace{-0.1em}
    \caption{Software implementation of \texttt{PACIA} instruction as a function.}
    \label{fig:paciaimplementation}
   \vspace{-0.1em}
  \end{figure}

Our implementation of \na uses a compile-time flag to indicate whether the
compiled binary should use software-emulated PAC or hardware-based PAC instructions.
Figure~\ref{fig:pacimplementation} shows an example of the inline assembly.

\begin{figure}
    \begin{lstlisting}
#if PACENABLED
    asm (
    "mov %
    :
    : "r" (ptr));
    asm(
    "pacia %
    : "=r" (ptr)
    : "r" (id));
#else
    ptr =__pacia(ptr,id);
#endif
    \end{lstlisting}
    \vspace{-0.1em}
    \caption{When the \texttt{PACENABLED} flag is enabled during the compile-time,
    actual PAC instructions are generated for the final binary. Otherwise, software
    implementation of the corresponding instructions is invoked. This implementation
    helps to test the design on an SoC that does not support the ARMv8.3 instruction set.}
    \label{fig:pacimplementation}
   \vspace{-0.1em}
  \end{figure}

\subsection{Security Evaluation}
\label{sec:evalsec}

The security evaluation is a functional test and serves two purposes: (1)
testing \na's compatibility with the underlying hardware feature, namely PAC,
and (2) testing \na's detection of temporal memory corruptions and its robustness
against evasions. Similar to the previous work~\cite{liljestrand2018pac}, we
chose the ARM FVP simulator for this functional test because no development
board with PAC extension exists at the moment and FVP includes ARM's official
and fully functional PAC simulation.

We performed the security evaluation using 150 C programs selected from the NIST
Juliet test suite~\cite{black2018juliet}. We chose the Juliet Suite for two
reasons. First, it is the largest of its kind and contains both vulnerable and
non-vulnerable versions of programs. The vulnerable programs, covering the
common types of temporal memory corruptions, are ideal for our security
evaluation. We used the non-vulnerable/patched counterparts for a compatibility
test. Second, unlike the generic CPU benchmarks, the test programs in Juliet
were made for security testing without being computationally demanding. They run
smoothly on FVP, which is a whole-system (slow) simulator and cannot run
computation-intensive programs without halting or crashing. Therefore, using the
Juliet programs allows us to focus on evaluating \na in terms of security while
avoiding high computation loads that FVP cannot handle. We conducted a separate
performance evaluation of \na (\S\ref{sec:evalperf}) using much demanding CPU
benchmarks.  

\begin{table}[t]
    \resizebox{0.49\textwidth}{!}{
        \begin{tabular}{l|cc|cc}
        \toprule
        Vulnerability                      & CWE Cat. & \# of Prog. &  \na & PAC / PARTS\cite{liljestrand2018pac} \\ \hhline{=====}
        Double-Free            & 415 & 50     & \pie{360}  & \pie{0}  \\ \hline
        Use-After-Free         & 416 & 50     & \pie{360} & \pie{0}  \\ \bottomrule
        Invalid-Free         & 761 & 50     & \pie{360} & \pie{0}  \\ \bottomrule
        \end{tabular}
        }
        \caption {Selection of 150 vulnerable programs from the Juliet Test Suite and detection results.}
        \label{table:testset}
\end{table}

The 150 Juliet tests include double-free, use-after-free and invalid-free bugs.
Table \ref{table:testset} shows the CWE (Common Weakness Enumeration) categories and
the number of Juliet tests selected in each vulnerability category. When running
with \na enabled, the vulnerable programs all terminated immediately before the
bugs were triggered. We also ran the non-vulnerable/patched version of the test
programs with \na enabled. All these programs finished properly without any
crash or halt. The result shows that \na achieved a 100\% detection accuracy and
did not cause any compatibility issues: it did not miss a single temporal
memory corruption in any category; it did not alert or crash the programs when
no temporal memory corruption was triggered. 

The right-half of Table \ref{table:testset} shows a comparison between \na and
PAC/PARTS~\cite{liljestrand2018pac}. PAC and PARTS were not designed
for detecting temporal memory corruptions and therefore cannot detect any. This
comparison underlines our novel use of PAC for addressing a critical security
vulnerability class, which is not considered or detectable by the original design of PAC or
previous work using PAC.

\begin{table}
    \resizebox{0.49\textwidth}{!}{
    \centering
    \begin{tabular}{l|ccccc}
    \hline\toprule
      Application     & CVE  & Vulnerability Type  & Detection   \\
    \hline
    libpng    &  CVE-2019-7317  & UAF & \ding{51} \\
    sqllite    &   CVE-2019-5018 & UAF  & \ding{51}\\
    curl  & CVE-2019-5481   & DF & \ding{51}\\
    libgd & CVE-2019-6978   & DF & \ding{51} \\
    
    \bottomrule\hline
    \end{tabular}
    }
    \caption{Effectiveness of \na for detecting real-world vulnerabilities.  }
    \label{table:cve}
\end{table}

\point{Case study of real-world vulnerabilities}
Besides the Juliet test programs, we also surveyed four recent temporal memory
corruption vulnerabilities in real software (Table \ref{table:cve}). Since FVP
cannot run these entire programs, we performed a manual analysis and verified
that \na can prevent the exploitations of all these vulnerabilities. For instance, 
To exploit CVE-2019-5481, an attacker sends a crafted request, which
\texttt{realloc()} fails to handle. On the exit path, the pointer is freed.
During the cleaning phase, the pointer is freed one more time. Since these two
steps are far apart, programmers can easily miss the bug. When \na is enabled, the
ID of the object is changed after the first free and thus causes an
authentication failure when the second free is about to happen. 
 
Take CVE-2019-7317 as another example, shown in Figure~\ref{fig:cve20197317}.
Line 5 indirectly calls \texttt{png\_image\_free\_function}, which frees the
memory referenced by \texttt{arg}, an alias of \texttt{image}. Later, Line 7
dereferences \texttt{image}, resulting in use-after-free. This bug can be
extremely difficult to discover either manually or using analysis tools, due to
the layers of function and object aliasing. \na handles aliasing naturally
thanks to its points-to authentication scheme. \na can catch this bug right
before it is triggered due to the \ac mismatch.

\begin{figure}[t]
    \begin{lstlisting}
if (result != 0)
{
  image->opaque->error_buf = safe_jmpbuf;
// calling png_image_free_function() indirectly
  result = function(arg);
}
image->opaque->error_buf = saved_error_buf;
    \end{lstlisting}
    \vspace{-0.1em}
    \caption{In CVE-2019-7317, the \texttt{png\_image\_free\_function} is called indirectly
    and the \textit{image} pointer is passed to it as an argument. In this case, the image 
    pointer is freed in the caller and then used in line 7, which is UAF error. }
    \label{fig:cve20197317}
   \vspace{-0.4em}
  \end{figure}

\point{Robustness evaluation}
We created a small set of programs that contain both temporal and spatial memory
corruptions to evaluate the robustness of \na. This scenario is analogous to the
real-world attacks where a powerful attacker can exploit arbitrary memory write and
temporal memory corruptions. We selected 30 programs from Juliet test suite in 3
different categories. Then, we injected memory overwrite vulnerabilities such as
buffer-overflow to them, which allow an attacker to overwrite the \na metadata. 
Such an attacker can bypass previous protections,
such as CETS, which simply compare the plain ID of the key and object. However, the attacker
cannot bypass \na because she cannot generate valid \ac without knowing the secret
PAC key. As expected, we triggered those vulnerabilities and \na detected all.

\subsection{Performance Evaluation}
\label{sec:evalperf}

We had to switch from the FVP simulator to a Raspberry Pi 4 (ARMv8-A Cortex A53)
for conducting the performance evaluation because the simulator could not run
computation-intensive benchmarks. Due to the lack of hardware support for PAC
on the Raspberry Pi SoC, we used our own software implementation of PAC in this
evaluation. This experiment provides an upper bound of the performance
overhead of \na (\ie the overhead should be lower on devices with hardware PAC
support).

\begin{figure}[t]
    \centering
    \includegraphics[scale = 0.55]{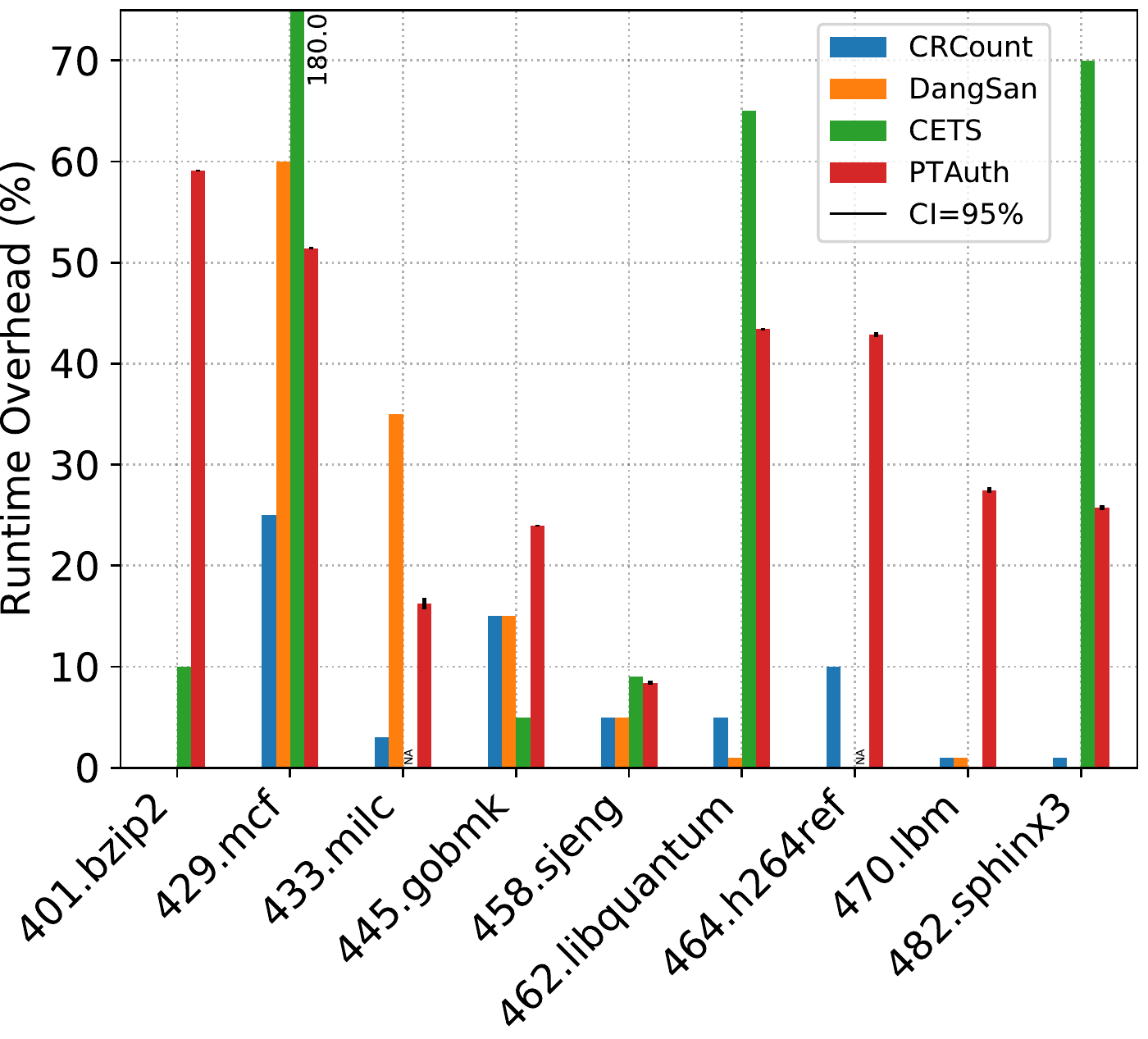}
    \caption{Runtime overhead (95\% confidence interval) on SPEC CPU2006 and comparison with CRCount, DangSan and CETS. }
    \label{fig:specoverhead}
\end{figure}

We tested \na on the SPEC CPU2006 benchmarks. They are appropriate for the
performance evaluation since they are memory- and CPU-intensive.
Figure~\ref{fig:specoverhead} shows the runtime overhead of \na with 95\%
confidence interval. The interval bars are barely visible in the figure due to
the relatively stable results.  
Figure~\ref{fig:boxplots} in Appendix~\ref{sec:app:distri} shows more clearly, for each benchmark, the concentrated distribution and the narrow standard deviation of the measured overhead. 
The overhead varies across different benchmarks
because the use of data pointers and dynamically
allocated objects in some benchmarks is more prevalent than in other benchmarks. For instance,
although \textit{mcf} is a small program, it uses many data pointers and
requires more \ac checks than other benchmarks.

We compared \na with the closely related works, including
CRCount~\cite{shin2019crcount}, DangSan~\cite{van2017dangsan}, and
CETS~\cite{nagarakatte2010cets}. These prior techniques are either based on
pointer invalidation or object (lock) invalidation. In our comparison, we skipped
DangNull~\cite{lee2015preventing} because DangSan outperforms it. Since the
source code of CRCount is not available and DangSan and CETS are not
compatible with the ARM architecture, we used the reported numbers in the
papers for comparison. 
The nine C benchmarks in Figure~\ref{fig:specoverhead} were selected because they are both compatible with our current
implementation and were used in the previous works. 
We note that the compared papers did not use the same set of benchmarks in their evaluation. 
Some of them did not report the performance numbers for all nine benchmarks. For example, 
CETS was not evaluated on {\tt 433.milc} and {\tt 464.h264ref}. 

The geometric mean overhead of
\na on all benchmarks is 26\%. The number around 5\% for CRCount,
1\% for DangSan, and 10\% for CETS.
Although \na
appears to incur much higher overhead than the others, we note that this
comparison is not completely fair because, unlike \na, the other systems
require external protection of their metadata (\eg bound checkers), which incurs additional overhead
not captured in this comparison. 

\begin{figure}[t]
    \centering
    \includegraphics[scale = 0.55]{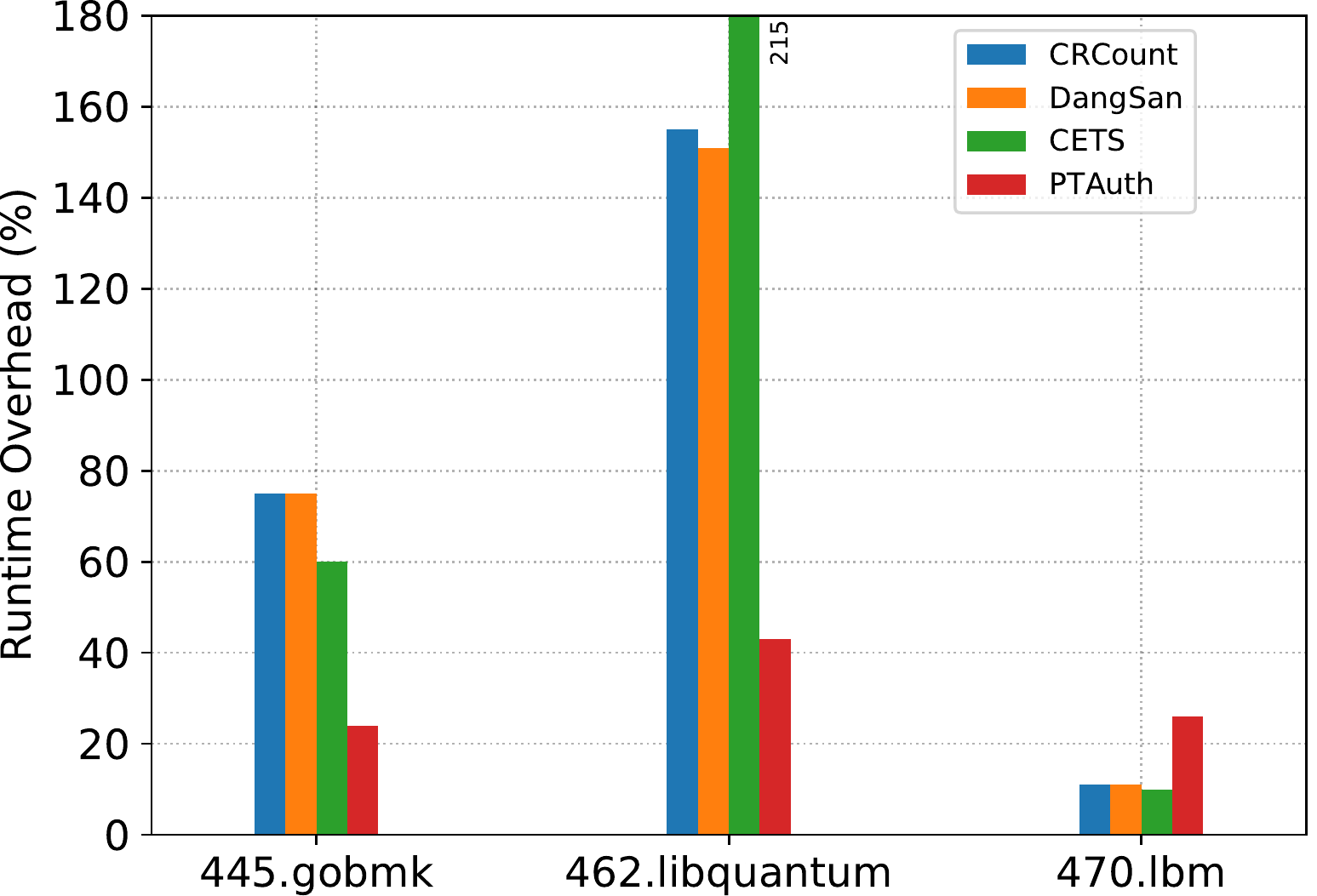}
    \caption{Runtime overhead of temporal memory corruption detectors when they are combined with SoftBound to protect their metadata. \na is a stand-alone system and does not need an external system to protect its metadata.  }
    \label{fig:softboundprotected}
\end{figure}

For this reason, we conducted another experiment, where we added the reported
runtime overhead of SoftBound~\cite{softbound} to the overhead of DangSan,
CRCount and CETS.  This combined overhead represents what these systems would
incur when they are deployed with the required external protection and made as
robust as \na. 
The results are shown in Figure~\ref{fig:softboundprotected}. Only three of the
nine benchmarks were tested in ~\cite{softbound} and thus were included in this
comparison. Clearly, \na incurs much less overhead than the other systems on 
two out of the three benchmarks. 

\point{Statistical significance}
To prove that our performance result is statistically significant, we perform
the following hypothesis testing and show that the performance overhead of \na
has a statistically significant upper bound at 42\% (with a P-value under
$0.05$). We construct the
\textit{Null Hypothesis} that ``the runtime overhead of \na is {\bf not}
below 42\%''. We show below that this Null Hypothesis can be {\bf rejected}.
We calculate the Z-score $= \frac{M - \mu}{\frac{\sigma}{\sqrt{n}}}$, where $M$
is the measured average runtime overhead of \na (i.e., 33.2\%), $\sigma$ is the
standard deviation of the measured overheads (i.e., 0.159), $n$ is the sample
size (i.e., 9), and  $\mu$ is the overhead bound stated in the Null Hypothesis
(i.e., 42\%). The calculated Z-score is $-1.66038$. Its corresponding P-value is
$0.04846$, which is below the widely accepted significance level of $0.05$.
The result of this hypothesis testing shows that a statistically significant
upper bound for \na's overhead can be established at 42\%.

\begin{figure}[t]
    \centering
    \includegraphics[scale = 0.55]{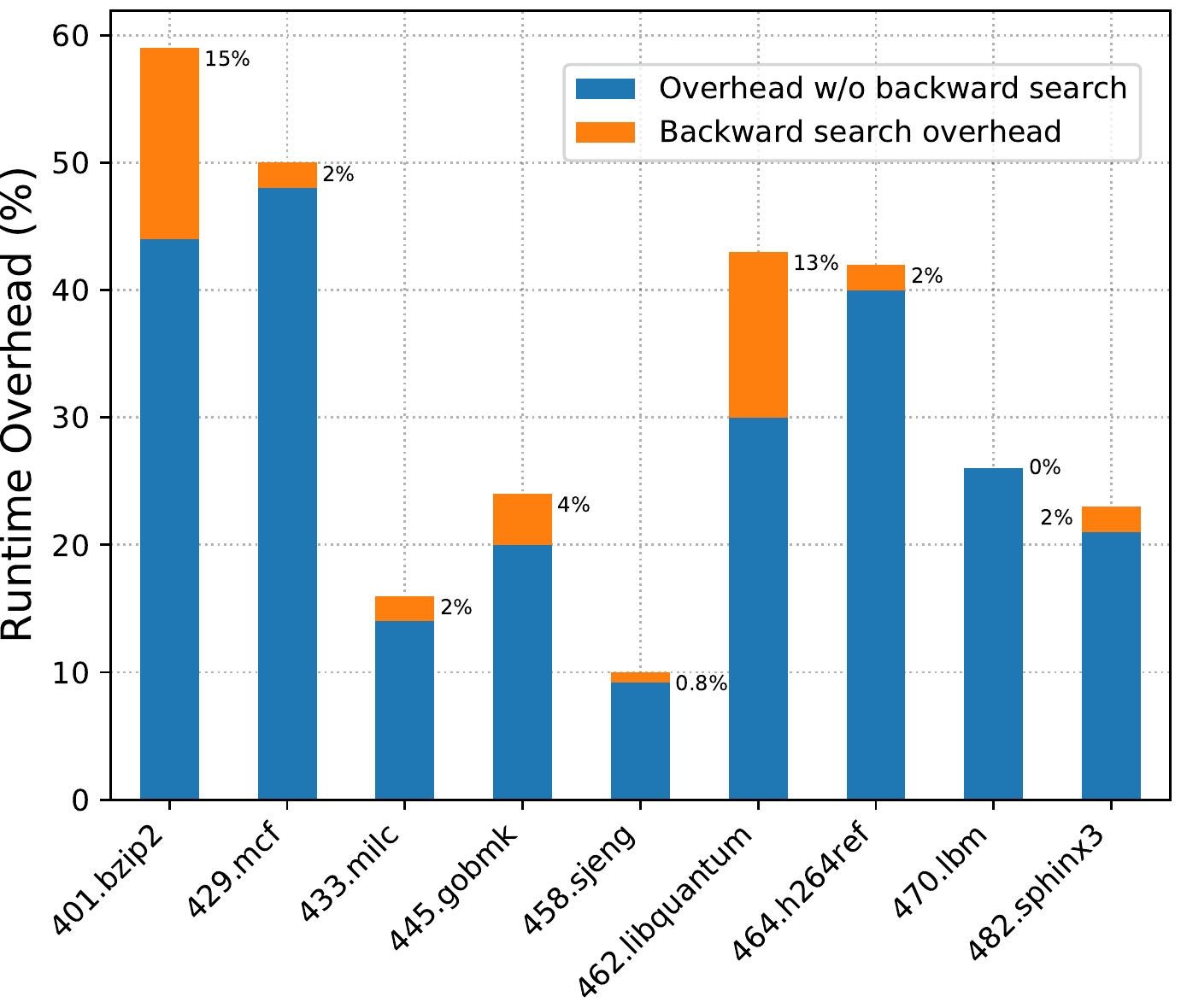}
    \caption{The overhead of backward search in each benchmark.}
    \label{fig:backscanning}
\end{figure}

\point{Backward search overhead}
Backward search incurs the worst-case runtime overhead when
many large memory objects exist with many sub-objects referenced
directly by pointers. However, in practice, we observed that this worst-case scenario is quite rare and the overhead of
backward search is generally low.  Figure~\ref{fig:backscanning} shows the overhead caused
by backward search in each benchmark, as part of the overall overhead. The main
reason for the low overhead of backward search is that most large objects are of
{\tt struct} type. The fields/sub-objects of the large objects are often
accessed via an index from the beginning of the large objects, rather than
direct or calculated pointers to the middle of the objects. Therefore, no
backward search is needed for those accesses to fields or sub-objects.  Also,
pointer arithmetics is not frequently used in the benchmarks and regular
programs. However, {\tt 401.bzip2} and {\tt 462.libquantum} contain
more pointer arithmetic operations than other benchmarks. Furthermore, our
optimization (\S\ref{sec:optmization}) removes some unnecessary checks on
pointers.

\begin{figure}[t]
    \centering
    \includegraphics[scale = 0.55]{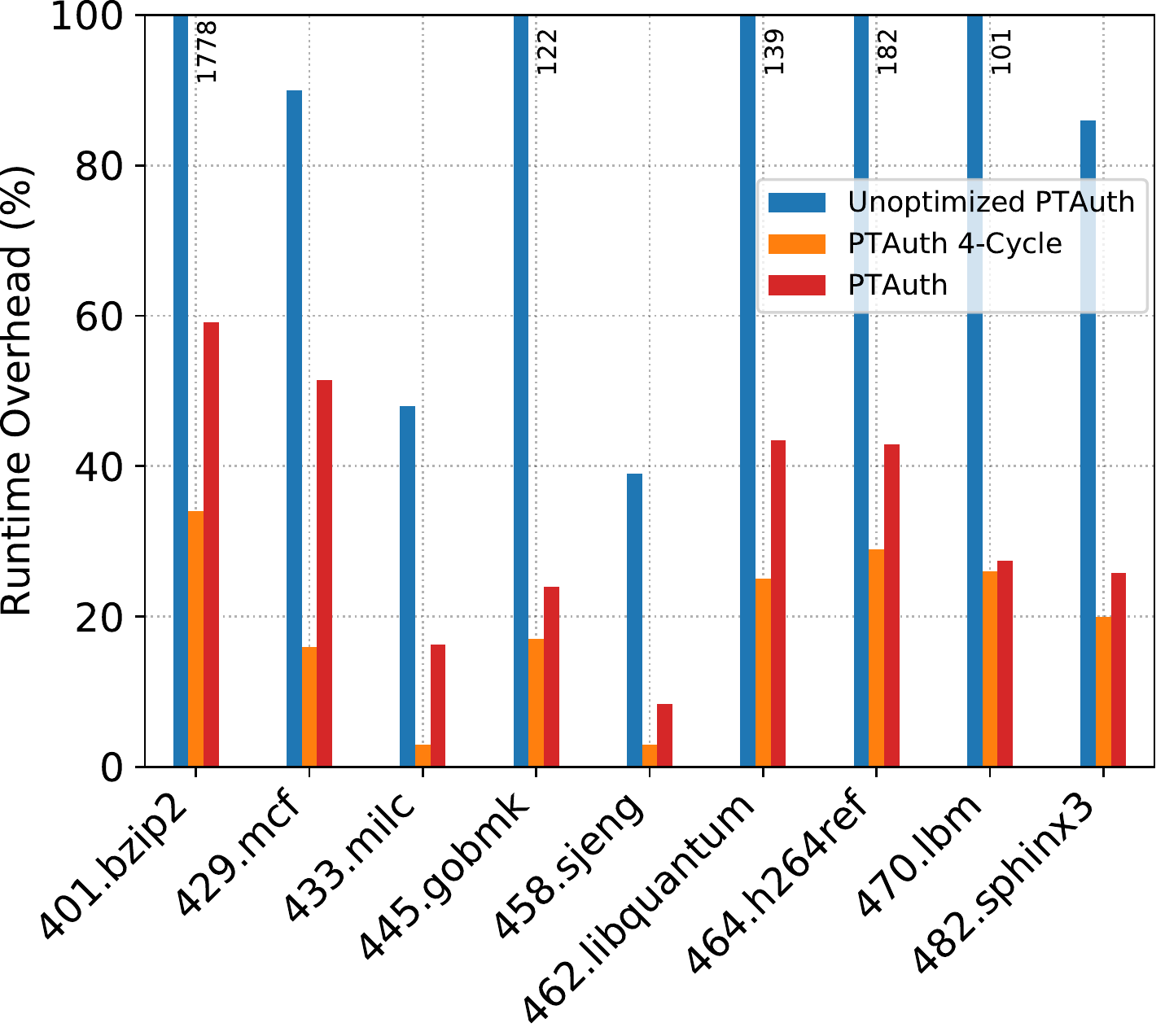}
    \caption{The overhead of \na with 4 CPU cycles and non-optimized implementation}
    \label{fig:ptauthversions}
\end{figure}

\point{Overhead under fixed-cycle PAC emulation}
In addition to evaluating \na's performance using our software implementation of
PAC, we conducted another experiment to estimate the performance of \na running
on future hardware with PAC support. This additional experiment was inspired by
PARTS~\cite{liljestrand2018pac}, where we used equal-cycle {\tt NOP}
instructions to replace PAC instructions (assuming each PAC instruction takes 4
CPU cycles as per~\cite{liljestrand2018pac}). Unlike PARTS, our system cannot
fully function when PAC instructions are simply replaced with NOPs. This is
because the backward search technique requires correct \ac produced by PAC
instructions. Therefore, in this experiment, we disabled the backward search and
the corresponding checks on sub-objects. This experiment is only meant to
complement the main performance evaluation (the one using the software-based
PAC). 
Figure~\ref{fig:ptauthversions} shows the runtime overhead on the benchmarks
assuming each PAC instruction takes exact 4 cycles. 
Base on this result, we expect the runtime overhead of \na to decrease to 20\% when \na runs on devices that support hardware PAC.

\point{Optimization benefit}  In order to measure the effectiveness of the
optimization described in \S\ref{sec:optmization}, we disabled the optimization
during the instrumentation and conducted the experiment again.
Figure~\ref{fig:ptauthversions} shows the benefits of the optimization. As
expected, the intra-procedural analysis for eliminating unnecessary checks
reduces, on average, 72.8\% of the overhead.

\point{Memory overhead}
In our design, pointer metadata is stored in the unused bits of a pointer and
requires no extra memory. The only source of \na's memory overhead is the extra
8-byte memory allocated for storing each object ID. We reduce this memory
overhead based on the following observation. The {\tt ptmalloc} allocator in the
\textit{glibc} of Linux appends extra paddings to objects when object sizes are
not 32-byte aligned. For instance, when a 16-byte object  is allocated, it is
padded to 32 bytes. For objects with such paddings,  \na makes use of the
padding bytes for storing object IDs, without requiring additional memory space.

\begin{figure}[t]
    \centering
    \includegraphics[scale = 0.55]{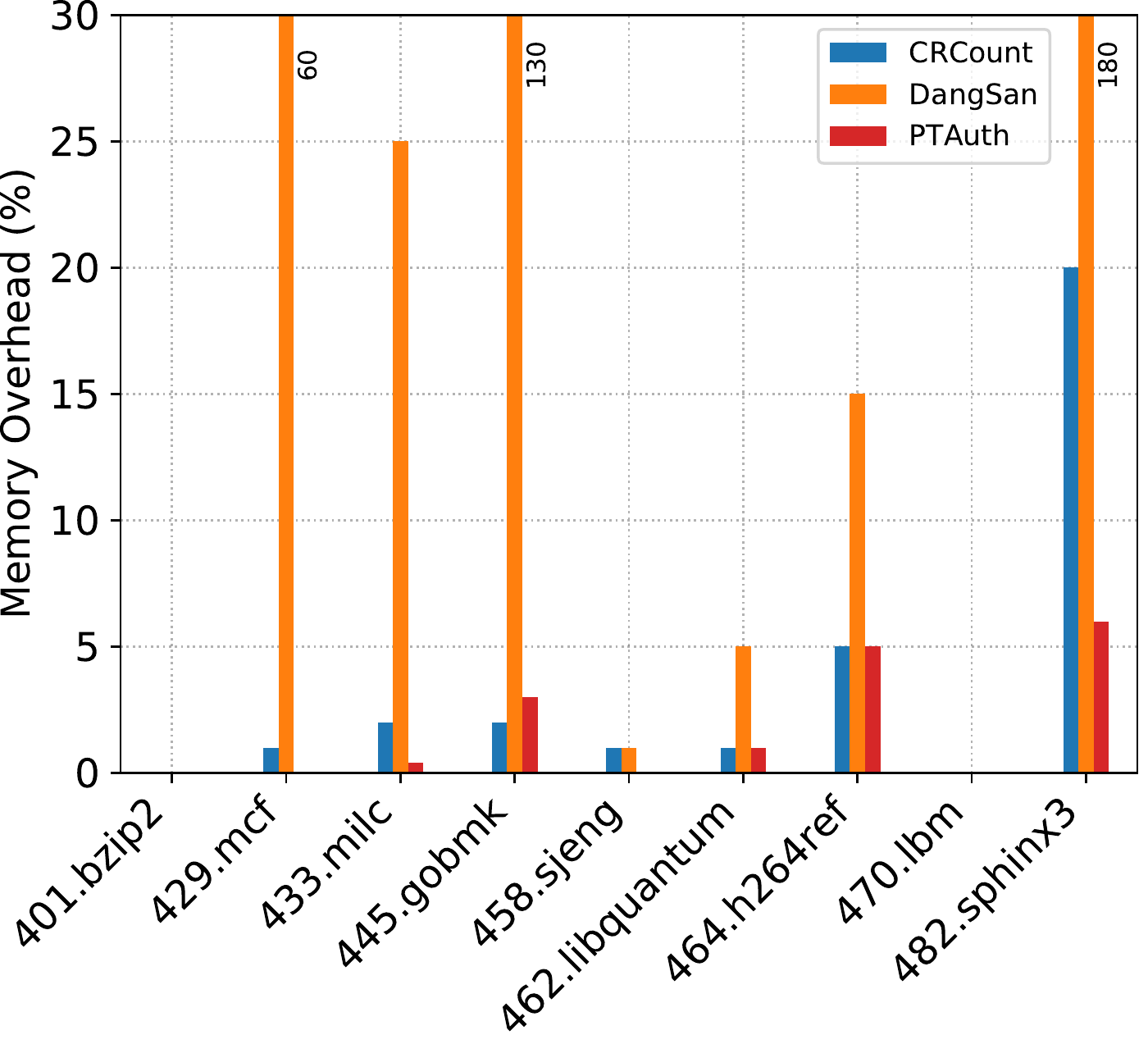}
    \caption{Memory overhead on SPEC CPU2006 and comparison with CRCount and DangSan. CETS have not reported any memory overhead. }
    \label{fig:specmemoryoverhead}
\end{figure}

To evaluate the memory overhead of \na, we measured the maximum resident set
size (Max RSS) of the instrumented SPEC CPU2006 benchmarks. 
Max RSS (\ie the peak physical memory allocation for a process) is the metric used by
related works. We adopted the same metric to perform a fair comparison. 
Figure~\ref{fig:specmemoryoverhead}
illustrates the memory overhead caused by \na, DangSan, and CRCount on each
benchmark. The geometric mean of \na's memory overhead is 2\%. This
number is 2\% for CRCount and 15\% for DangSan. CETS did not report memory
overhead. 

In addition to maximum RSS, we also measured the memory overhead of \na in terms
of mean RSS, a metric that captures the memory usage throughout the entire
process execution, as opposed to the peak usage at one moment. We calculated mean RSS
by taking RSS every five seconds during the lifetime of a process and then computing the mean.
The mean RSS overhead of \na is 1\%. The related works
did not report this number and thus cannot be compared with in this regard.

\section{Discussion and Limitation}
\label{sec:discuss}

\point{Multi-threading} 
Similar to previous works, the current \na prototype does not support
multi-threaded programs, mostly due to implementation-level simplifications. To
make \na work on multi-threaded programs, each memory (de)allocation and the
resulting metadata updating operation need to be atomic, or the metadata
protected by a lock. Without the atomicity or synchronization, \na's metatdata
may become stale or invalid when a race condition occurs, leading to missed or
falsely detected temporal memory errors. It is worth noting that the unnecessary
check removal (one optimization discussed in \S\ref{sec:optmization}) is not
threading-safe and needs to be disabled on multi-threaded programs.

\point{Stack use-after-free}
Since the deallocation of stack objects is implicitly triggered by function
returns, double-free bugs cannot happen to stack objects. However,
use-after-free bugs on stack objects, though uncommon, may happen when the
address of a stack object is taken and stored in a global variable that is later
mistakenly freed. 
The current design of \na is focused on detecting heap-based temporal errors,
which are more prevalent and critical than stack-based use-after-free.
In theory, \na can be extended to detect the latter as follows. 

We refer to stack objects referenced by global pointers as address-taken objects
(\ie stack objects potentially vulnerable to use-after-free). Since object
allocation and deallocation on the stack are different from those on the heap,
protecting address-taken objects require special treatment. First, \na needs to
identify address-taken objects in stacks via a simple intra-procedural data-flow
analysis. Then, \na needs to allocate extra 8 bytes at the beginning of each
address-taken object. This extra header is used for storing the metadata, which
is initialized upon the creation of the stack frame (\ie in the prologue of the
corresponding function). \na also needs to instrument address-taken operations
on stack objects to generate \ac for the resulting pointers. The authentication
scheme for address-taken objects on stacks and their pointers is the same as the
scheme for heap objects and pointers.    
Finally, \na needs to invalidate all metadata of address-taken objects in a function
epilogue, similar to what it does upon heap object deallocations. 

\point{PAC instructions}
In the current implementation, we have used both PAC instructions and the
software emulated implementation of the instructions. We used the FVP simulator to
run the PAC instruction. However, FVP is not a performance aware simulator. It
does not model cycle timing and all the instructions are executed in one
processor clock cycle~\cite{fvplimitation}. We also observed that large
benchmarks halt the FVP which prevented us from running performance experiments on it.
Since A12 Soc is proprietary and there is no public SoC available to test the
implementation, the reported runtime overhead is anticipated to be different in
real hardware. In other words, the actual PAC instructions are expected to be
faster than the software emulated instructions.

We leave these limitations for future work when the real hardware is available.

\section{Related work} 

\point{Safe C}
Memory corruption bugs are highly diverse and commonly targeted by software
attacks~\cite{szekeres2013sok}. Prior work introduced memory safety to the C
language via a safe type 
system~\cite{jim2002cyclone,necula2005ccured,dhurjati2006safecode,simpson2013memsafe}.
These safe languages are immune to temporal
vulnerabilities. However, they either impose a significant amount of memory and
runtime overhead or they are not applicable to protect legacy C/C++ codes. For
instance, Cyclone~\cite{jim2002cyclone} is a safe dialect of C which is is not
applicable to protect legacy codes. It is no longer supported but several ideas
of Cyclone have been implemented in Rust~\cite{rust,cyclonesite}. CCured needs
some annotations by the programmer. It also uses fat pointers to store metadata
which breaks the application binary interface (ABI).

\point{Safe memory allocator}
These systems prevent allocated objects from ending up at the same address of
freed objects
~\cite{silvestro2017freeguard,akritidis2010cling,berger2006diehard,novark2010dieharder}.
For instance, DieHard~\cite{berger2006diehard} and
DieHarder~\cite{novark2010dieharder} randomize the locations of allocated
objects in the heap and consequently provides probabilistic temporal memory
safety (\ie making object reuse or replacement difficult).
PartitionAlloc~\cite{partitionalloc} and Internet Explorer isolated
heap~\cite{isolatedheap} allocators prevent memory reuse by allocating objects
of different types or sizes in separate buckets. Although these schemes have low
runtime overhead, it has been shown that they can be bypassed on targeted
attacks~\cite{hariri2015abusing,isolatedheapbypass}. Moreover, they suffer from
a huge memory overhead caused by memory fragmentation.

\point{Memory error detectors}
Memory error
detectors~\cite{serebryany2012addresssanitizer,bruening2011practical,nethercote2007valgrind} 
are widely used among developers. However, due to the
high overhead, they are only suitable for debugging or non-production use.
AddressSanitizer~\cite{serebryany2012addresssanitizer} is a memory error
detector that creates shadow memory and red zones around objects. It detects
out-of-bounds accesses in heap, stack, and global objects, as well as use-after-free
bugs. However, it provides a probabilistic detection system for use-after-free
bugs which is susceptible to bypass~\cite{wimberley2013bypassing}.

\point{Pointer invalidation}
Another line of work focused on pointer invalidation.
DangNull~\cite{lee2015preventing}, DangSan~\cite{van2017dangsan},
FreeSentry~\cite{younan2015freesentry} and pSweeper~\cite{liu2018robust}
explicitly invalidate all the pointers to an object when the lifetime of the
object is finished. CRCount~\cite{shin2019crcount} uses a reference counting
approach for counting the number of pointers to an object. When there is no pointer to
an object, then it is freed. In this approach, the pointers are invalidated
implicitly during the runtime of the program. This approach suffers from memory
leak issue since some pointers are never invalidated. Consequently, the objects will
reside in memory for a long time. In general, pointer invalidation systems
need to keep a huge amount of metadata in the memory to track the relationship
between pointers and objects. Inevitably, those metadata are prone to corruption.

\point{Pointer dereference validation}
Some other approaches similar to our design, detect and prevent temporal corruption bugs by
pointer dereference
validation~\cite{nagarakatte2010cets,yong2003protecting,burow2018cup}.
CETS~\cite{nagarakatte2010cets} provides temporal safety by assigning a unique
identifier to each object and its pointers. The main challenge in this scheme is
that extra metadata for the pointers should be stored in the memory. Also, a
unique identifier should be assigned to each object and its pointers. Since
these metadata are stored disjointly, obtaining these information efficiently
during the runtime is challenging. In order to tackle this problem, in our design, we proposed an
inline metadata scheme for both pointers and objects. However, inline metadata
is prone to corruption by linear overflow. To address this problem, we used PAC
to guarantee the integrity of the metadata before using them. To sum up, our
approach reduces the high look-up table costs for loading the metadata and
provides integrity of the metadata in a unified design.

\point{Hardware-assisted schemes}
Similar to \na, there are some approaches that take advantage of hardware to
provide temporal safety. Oscar~\cite{dang2017oscar}, which is the following work
of~\cite{dhurjati2006efficiently}, is a page permission-based scheme to prevent
temporal memory safety violations in the heap. Basically, Oscar improves the
original idea of allocating each object in a separate page (similar to PageHeap
and Electric Fence~\cite{pageheap,electricfence}) to prevent UAF
vulnerabilities.

Another line of work relies on hardware to provide spatial and temporal
protections. Hardware-assisted AddressSanitizer
(HWASAN)~\cite{serebryany2018memory,hwasan} is the following work of
AddressSanitizer. HWASAN uses address tagging feature~\cite{addresstagging} to
implement a memory safety tool, similar to AddressSanitizer. Memory Tagging
Extension (MTE)~\cite{armv85a} has been introduced in ARMv8.5 for providing
spatial and temporal safety. However, the hardware is not available yet. Intel
MPX~\cite{OleksenkoMPX} was introduced by Intel to provide spatial safety.
However, due to the high-overhead, it was discontinued by the maintainers.

\section{Conclusion}
\label{sec:conc}

We presented a resilient and efficient points-to authentication scheme called \na, for
detecting temporal memory corruptions. 
By defining the
authentication codes (\ac) for pointers, our scheme allows for
convenient and simultaneous checking of metadata integrity and identities when
they are being accessed. The unified verification of the two properties
(integrity and identity) enables the unified detection of all kinds of temporal memory
corruptions in the heap.
\na uses PAC on ARMv8.3-A as a basic encryption/signing primitive during \ac
calculation, which is fast and secure thanks to the hardware-level support. \na
contains: (\rom{1}) a customized compiler for instrumenting programs with
necessary inline checks, (\rom{2}) a runtime library for \ac generation and
authentication, and (\rom{3}) a set of OS patches for PAC-related CPU
configuration. Our evaluation on 150 vulnerable programs shows that \na
detects all 3 categories of temporal memory corruptions
with a runtime overhead of 26\% (using software-based PAC) and 2\% memory
overhead.

\section*{Acknowledgment}
The authors would like to thank the anonymous reviewers for their help with the revision of this paper. This project was supported by 
the Office of Naval Research (Grant\#: N00014-18-1-2043 and N00014-17-1-2891) %
and the Army Research Office (Grant\#: W911NF-18-1-0093).
Any opinions, findings, and conclusions or recommendations expressed in this
paper are those of the authors and do not necessarily reflect the views of the
funding agencies.

\bibliographystyle{plain}
\bibliography{paper}

\section*{}
\label{sec:appendix}

\begin{appendices}

\section{Distribution of Runtime Overheads}
\label{sec:app:distri}

\begin{figure}[ht]
\centering
\includegraphics[scale=0.45]{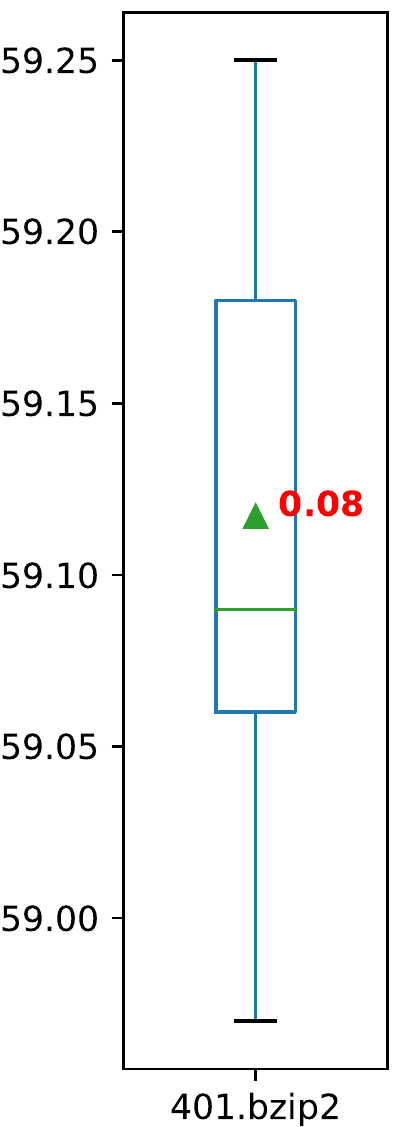}
\includegraphics[scale=0.45]{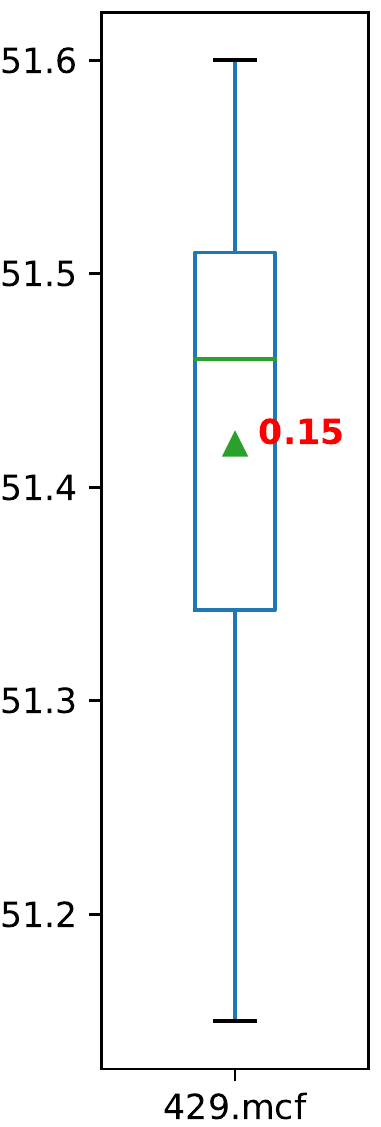}
\includegraphics[scale=0.45]{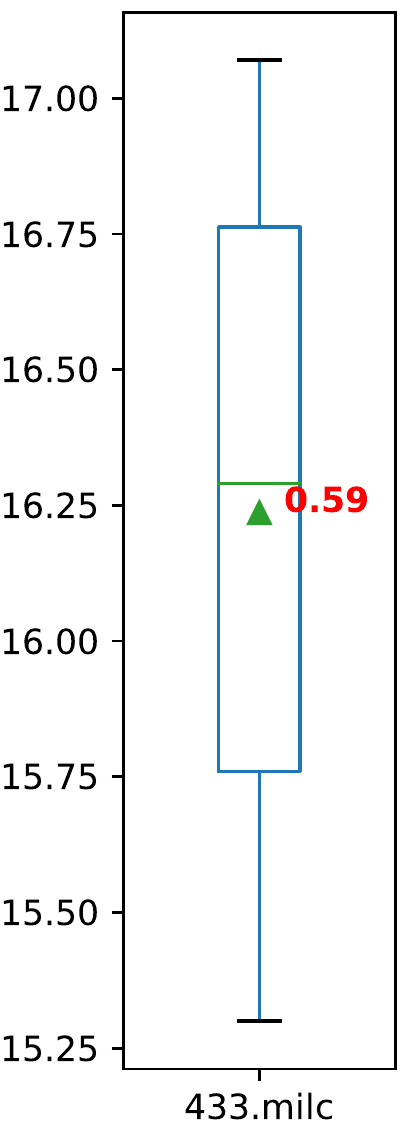}
\includegraphics[scale=0.45]{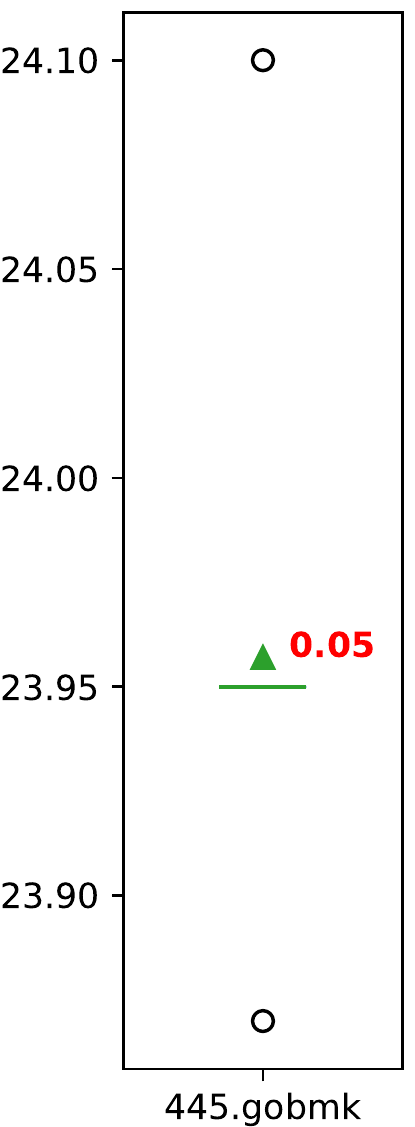}
\includegraphics[scale=0.45]{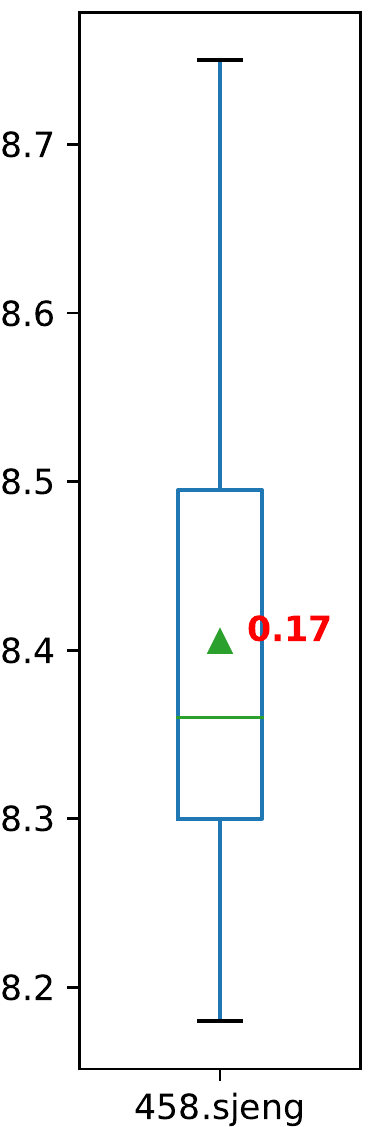}
\includegraphics[scale=0.45]{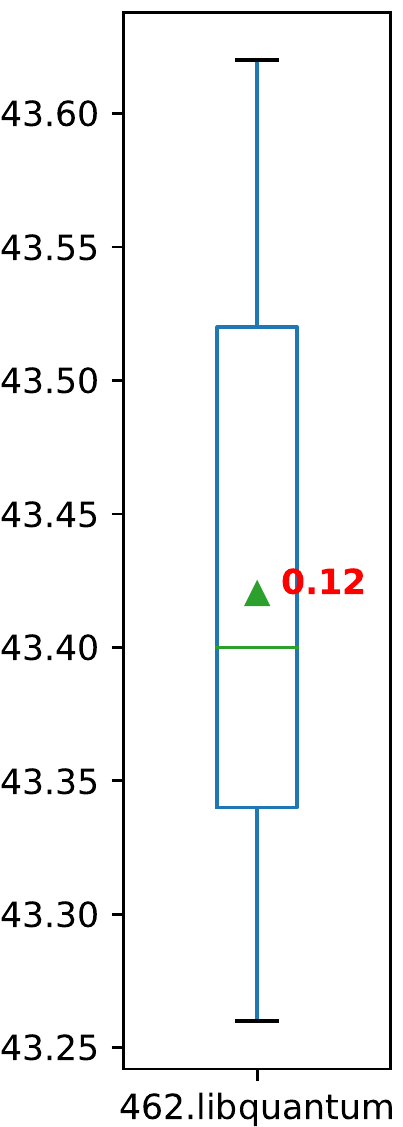}
\includegraphics[scale=0.45]{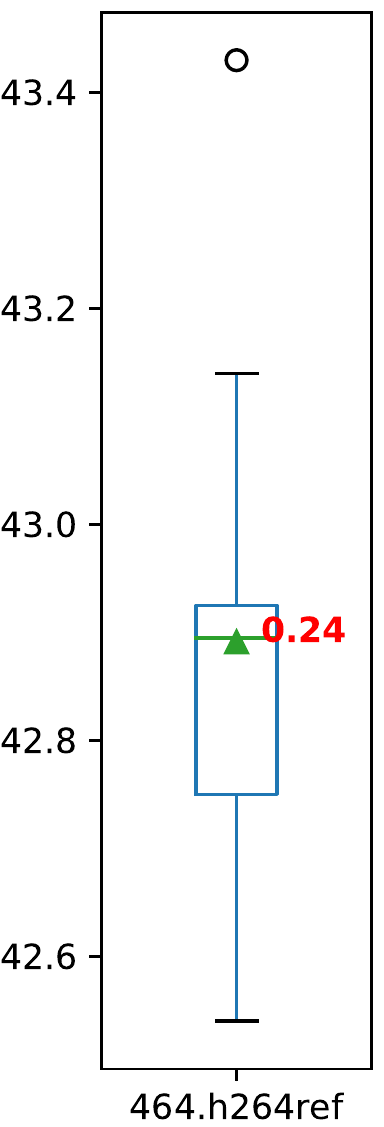}
\includegraphics[scale=0.45]{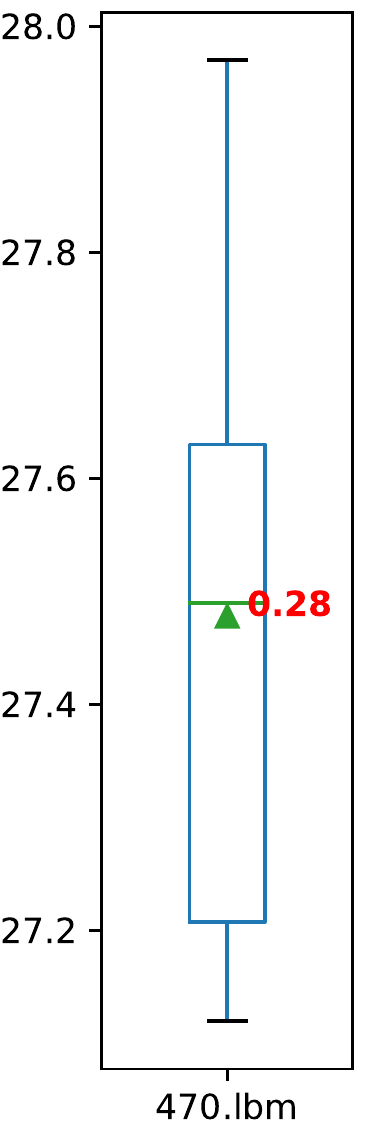}
\includegraphics[scale=0.45]{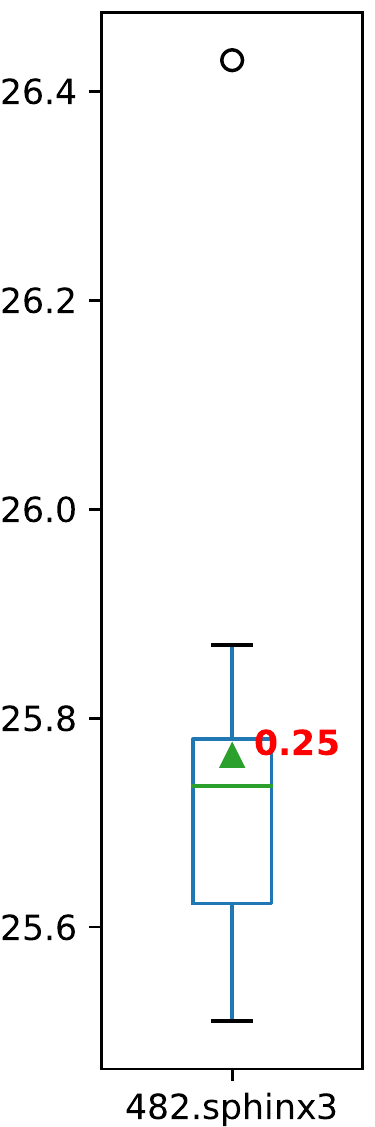}
\caption{Distribution of \na's runtime overhead on different benchmarks. Per-benchmark scales are used to clearly show the overhead distribution on each individual benchmarks.}
\label{fig:boxplots}
\end{figure}

Figure~\ref{fig:boxplots} shows the distribution of runtime overheads obtained
by running \na 10 times on all the benchmarks. The green triangles indicate the
mean and the red numbers are the standard deviation. Note that the box plots do
not use the same scale because the overhead varies significantly across the
benchmarks. The standard deviation on all benchmarks are fairly low (i.e., less
than 0.6\%), indicating that the overhead values distribute closely around the
mean. This result confirms that the overhead result is reliable.

\end{appendices}

\end{document}